\documentclass[amsmath,amssymb,prl,aps,superscriptaddress,twocolumn]{revtex4-1}
\pdfoutput=1
\usepackage{amsmath,amsfonts,amssymb,amsthm,graphics,graphicx,epsfig,bbm}

\usepackage[colorlinks=true,citecolor=blue,linkcolor=blue,urlcolor=blue]{hyperref}

\usepackage{subfigure}

\usepackage{dcolumn}
\usepackage{bm}
\usepackage{color}
\usepackage{epstopdf}

\usepackage{amstext}
\usepackage{latexsym}

\usepackage{psfrag}
\usepackage{xcolor}
\usepackage[normalem]{ulem}
\usepackage{dsfont}
\usepackage{txfonts}
\usepackage{cases}
\usepackage{ifthen}
\usepackage{algorithm}
\usepackage{algpseudocode}
\usepackage{soul}

\graphicspath{{./Images/},{./Imagesappendix/}}

\newcommand{\ket}[1]{\ensuremath{\left\vert #1 \right\rangle}}
\newcommand{\bra}[1]{\ensuremath{\left\langle #1 \right\vert}}

\newcommand{\braket}[2]{\langle #1 \vert #2 \rangle}

\newcommand{\mean}[1]{\left\langle #1 \right\rangle}

\newcommand{\an}[2]{\ensuremath{\hat{#1}^{\protect\phantom{\dagger}}_{#2}}}
\newcommand{\cn}[2]{\ensuremath{\hat{#1}^\dagger_{#2}}}

\newcommand{\expU}[1]{\ensuremath{e^{#1}}}
\newcommand{\abs}[1]{\left|#1\right|}

\newcommand{\pdif}[2]{\ensuremath{\frac{\partial#1}{\partial#2}}}

\newcommand{\subfigimg}[3][,]{%
	\setbox1=\hbox{\includegraphics[#1]{#3}}
	\leavevmode\rlap{\usebox1}
	\rlap{\hspace*{2pt}\raisebox{\dimexpr\ht1-0.5\baselineskip}{{\bfseries \large\textsf{#2}}}}
	\phantom{\usebox1}
}

\newcommand{\idg}[1]{{\bfseries #1)}}

\begin{document}
	
\title{Classifying global state preparation via deep reinforcement learning}






\author{Tobias Haug}
\affiliation{Centre for Quantum Technologies, National University of Singapore,
3 Science Drive 2, Singapore 117543, Singapore}
\author{Wai-Keong Mok}
\affiliation{Centre for Quantum Technologies, National University of Singapore,
	3 Science Drive 2, Singapore 117543, Singapore}
\affiliation{Department of Electronics and Photonics, Institute of High Performance Computing, 1 Fusionopolis Way, 16-16 Connexis,
Singapore 138632, Singapore}
\author{Jia-Bin You}
\affiliation{Department of Electronics and Photonics, Institute of High Performance Computing, 1 Fusionopolis Way, 16-16 Connexis,
Singapore 138632, Singapore}
\author{Wenzu Zhang}
\affiliation{Department of Electronics and Photonics, Institute of High Performance Computing, 1 Fusionopolis Way, 16-16 Connexis,
Singapore 138632, Singapore}
\author{Ching Eng Png}
\affiliation{Department of Electronics and Photonics, Institute of High Performance Computing, 1 Fusionopolis Way, 16-16 Connexis,
Singapore 138632, Singapore}
\author{Leong-Chuan Kwek}
\affiliation{Centre for Quantum Technologies, National University of Singapore,
3 Science Drive 2, Singapore 117543, Singapore}
\affiliation{MajuLab, CNRS-UNS-NUS-NTU International Joint Research Unit, UMI 3654, Singapore}
\affiliation{National Institute of Education and Institute of Advanced Studies,
Nanyang Technological University, 1 Nanyang Walk, Singapore 637616}
\affiliation{School of Electrical and Electronic Engineering Block S2.1, 50 Nanyang Avenue,
Singapore 639798 }

\date{\today}

\begin{abstract}
Quantum information processing often requires the preparation of arbitrary quantum states, such as all the states on the Bloch sphere for two-level systems. While numerical optimization can prepare individual target states, they lack the ability to find general solutions that work for a large class of states in more complicated quantum systems.
Here, we demonstrate global quantum control by preparing a continuous set of states with deep reinforcement learning. The protocols are represented using neural networks, which automatically groups the protocols into similar types, which could be useful for finding classes of protocols and extracting physical insights.
As application, we generate arbitrary superposition states for the electron spin in complex multi-level nitrogen-vacancy centers, revealing classes of protocols characterized by specific preparation timescales. 
Our method could help improve control of near-term quantum computers, quantum sensing devices and quantum simulations.
\end{abstract}

\maketitle

\section{Introduction}
Deep reinforcement learning has revolutionized computer control over complex games \cite{schulman2015high,mnih2016asynchronous,silver2016mastering}, which are notoriously difficult to optimize with established methods \cite{silver2016mastering}. 
Recently, reinforcement learning has also been successfully applied to a wide array of physics problems \cite{chen2013fidelity,bukov2018reinforcement,zhang2019does,bharti2019teach,haug2019engineering,dalgaard2020global,an2019deep,RevModPhys.91.045002,niu2019universal,porotti2019coherent,xu2019generalizable,bharti2020machine}.
A particular promising application is optimizing quantum control problems \cite{arrazola2019machine,niu2019universal,an2019deep,haug2019engineering,zhang2019does,dalgaard2020global}, which are of utmost importance to enable quantum technologies and quantum devices for information processing and quantum computation. 
Full control of a quantum system requires mastering a large amount of different protocols.  This task is key to efficiently manipulate quantum sensing devices and quantum computers.  For example, to drive an initial state to any arbitrary quantum state in a two-level system, a two-dimensional set of protocols has to be learned. Of course, for specific types of driving the parameterization is well-known  (e.g. in terms of rotations around the Bloch sphere) and there is a simple understanding of how to control the spin dynamics on the Bloch sphere.
However, for more complicated quantum systems or constrained driving parameters, no general description is available, which is a limiting factor in the control of quantum devices.  Standard numerical tools available can find control protocols for individual target quantum states, however it is difficult to find a class of protocols that can parameterize all the protocols as it can be highly disordered with many near-optimal solutions \cite{bukov2018reinforcement}.
Thus, we ask: how can one find classes of solutions to generate arbitrary states in more complicated models? And can they be used to extract physical insights?

\begin{figure*}[htbp]
	\centering
	\subfigimg[width=0.8\textwidth]{}{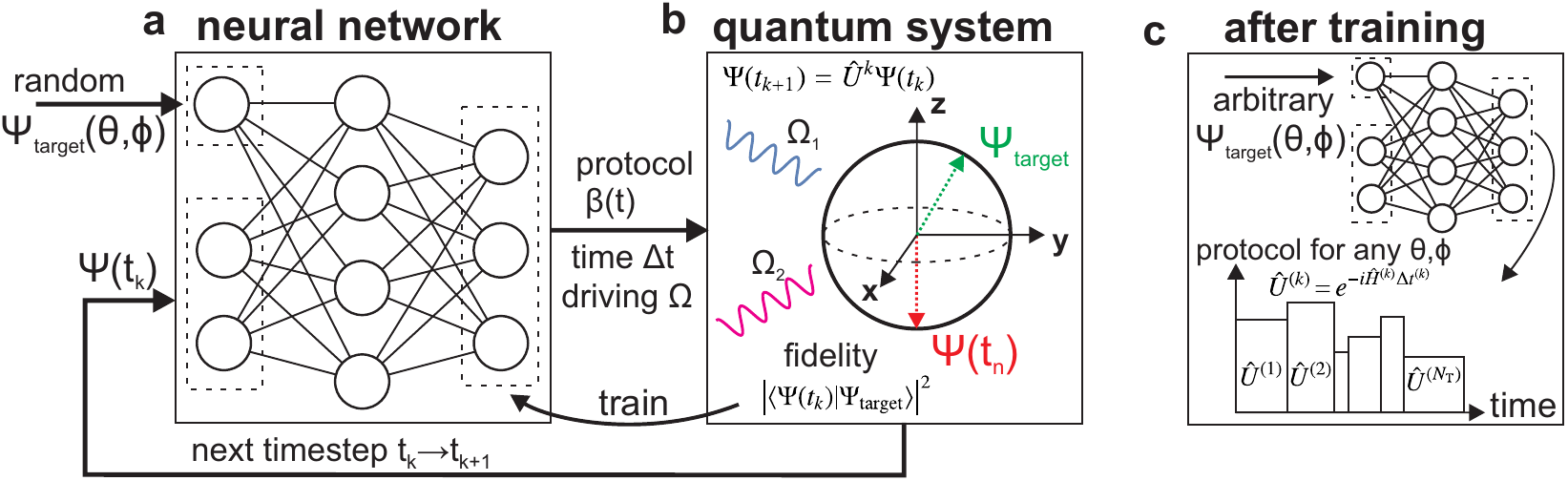}
	\caption{Overview of deep reinforcement algorithm to learn protocols to generate any target states. \idg{a,b} Learning scheme to generate piece-wise constant control protocol: At the first iteration step, a random target state $\Psi_\text{target}$ (parameterized by an angle) and the initial state $\Psi(t_0)$ is chosen. At iteration step $n+1$ of the algorithm, the neural network takes in $\Psi_\text{target}$ and the current wavefunction $\Psi(t_n)$. The output of the neural network generates the driving parameters (timestep $\Delta t$, driving strength $\Omega$) to manipulate the quantum system to reach $\Psi(t_{n+1})$, which is fed back to the neural network. These steps are repeated until final step $N_\text{T}$. This algorithm is repeated over many epochs to train the neural network to maximize the fidelity with the target state. \idg{c} After the training, the neural network returns optimized piece-wise constant control protocols to reach any target state $\Psi_\text{target}$.}
	\label{Sketch}
\end{figure*}


Here, instead of finding the optimal driving protocol for a single state, we propose a scalable method to learn all the driving protocols for global state preparation (over the continuous two-dimensional subspace represented by the Bloch sphere, embedded in a higher-dimensional Hilbert space) using deep reinforcement learning and parameterizing the protocols with neural networks. 
We discover that this approach automatically finds clusters of similar protocols. For example, it groups protocols according to how much time is needed to generate specific target states, which could be used to identify patterns and physical constraints in the protocols. For this multi-dimensional problem, conventional optimization often finds uncorrelated protocols in the target space, especially if there are multiple distinct protocols achieving similar fidelity. This makes it very difficult to interpolate between the protocols. Using our approach, the clustering of similar protocols is a consequence of effective interpolation by the neural network of nearby protocols within the same cluster, thereby achieving effective arbitrary state preparation.
The neural network is trained with random target states as input. This makes it naturally suited for parallelization and could allow it to scale to higher dimensions. In essence, our proposal shows a path to solve the key limitations of single-state learning or a transfer-learning approach as shown in \cite{niu2019universal}.

As a demonstration of our method, we apply it to control the electron spin in multi-level nitrogen-vacancy (NV) centers \cite{yale2013all,zhou2017accelerated,yale2016optical,tian2019optimal}. The multiple levels give rise to complicated coherent and dissipative dynamics, thus ruling out simple driving protocols such as those in two-level systems (e.g. quantum dot, transmon) and further increasing the difficulty of optimization.
The electron spin triplet ground states are characterized by a long lifetime which makes them ideal candidates for solid-state qubits used in quantum information processing \cite{gruber1997scanning,balasubramanian2009ultralong,maurer2012room}. However these states cannot be coupled directly using lasers, which would be important for many applications \cite{chu2015quantum,wang2014all}. 
Optical control can instead be achieved by indirect optical driving via a manifold of excited states, which renders finding fast control protocols a difficult problem. 
With our algorithm we find protocols to prepare arbitrary superposition states with a preparation speed of about half a nanosecond, which is much faster than protocols based on STIRAP control \cite{zhou2017accelerated} and reduces the impact of dissipation on the state preparation. Additionally, our protocols require only 9 steps, which is considerably less than in comparable approaches \cite{tian2019optimal,porotti2019coherent}. More importantly, our algorithm finds near-optimal protocols that are grouped into distinct classes. Each class spans a part of the Bloch sphere, and is characterized by similar driving protocols and the time needed to the prepare the quantum states.

\section{Learn global control protocols}
We now outline the algorithm to generate control protocols $\beta_{\theta,\phi}(t)$ to create all possible target states $\Psi_\text{target}(\theta,\phi)$ parametrized by angles $\theta$ and $\phi$ (see Fig.~\ref{Sketch}). The protocol is a piece-wise constant function of $N_\text{T}$ steps. 
At time $t_k$ the input to the neural network is the randomly sampled target state and a description of the system state (e.g. the current wavefunction $\Psi(t_k)$). The neural network output determines the parameters (driving strength $\Omega^{(k)}$, timestep $\Delta t^{(k)}$) for the piece-wise constant protocol used to evolve the wavefunction to the new state $\Psi(t_{k+1})=\hat{U}^{(k)}\Psi(t_k)$ with unitary $\hat{U}^{(k)}$ given by $\hat{U}^{(k)}=\exp(-i\hat{H}(\Omega^{(k)})\Delta t^{(k)})$.   
This new wavefunction is then input to the neural network to determine the next protocol step. This is repeated until the final timestep $N_\text{T}$ with total time $T=\sum_k \Delta t^{(k)}$, which concludes one training episode.
The goal is to drive the system such that wavefunction is as close as possible to a target state $\Psi_\text{target}$, measured by maximizing the fidelity
\begin{equation}\label{Fidelity} F=\abs{\braket{\Psi(t_{N_\text{T}})}{{\Psi_\text{target}}}}^2
\end{equation}
as a reward function. With each training episode, the neural network is trained with randomly sampled target states $\Psi_\text{target}(\theta,\phi)$ as input, until it converges and learns to represent protocols for all possible target states.  The neural network is trained using Proximal Policy optimization (PPO) \cite{schulman2017proximal}. The neural network is composed of two parts: The critic estimates at every step the final quality of the protocol to optimize itself, while the actor at each step returns a part of the protocol. We use the PPO algorithm implemented by the OpenAI Spinning Up library~\cite{spinningup}.

\section{Electron spin control in NV center}
\begin{figure*}[htbp]
	\centering
	\subfigimg[width=0.9\textwidth]{}{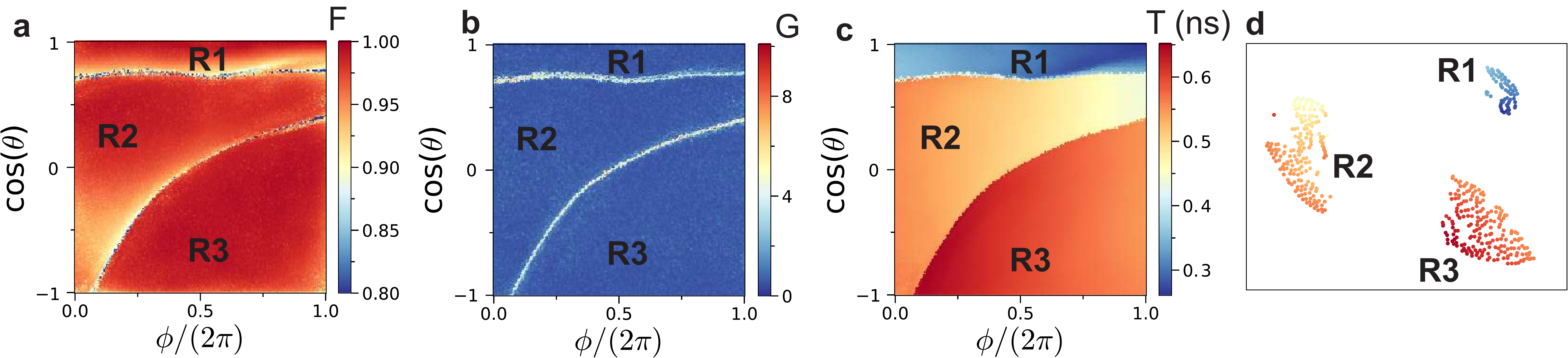}
	\caption{Create arbitrary quantum state $\Psi(\theta,\phi)$ (Eq.\ref{NVstate}, state parameterized by angles $\theta$, $\phi$) using deep reinforcement learning.  \idg{a} Fidelity $F$ (Eq.\ref{Fidelity}) of preparing state  (Mean fidelity $\langle F \rangle=0.972$) \idg{b} Protocol gradient $G(\theta,\phi)$ (Eq.\ref{grad}). Protocols are grouped into three distinct areas of low gradient (areas numbered with R1 to R3), divided by a sharp lines of large gradient.  \idg{c} protocol time $T$. The three areas are characterized by different protocol times $T$. 
		\idg{d} Two-dimensional representation of the distance  between protocols (determined by Eq.~\ref{dist}) using t-SNE algorithm. Color indicates protocol time. Protocols are close to each other if they are similar. Corresponding regions to the gradient plot in b) are marked with $R$.
		Parameters: $N_\text{T}=9$ timesteps, variable time per step with maximal time $0.2\text{ns}<T<0.8\text{ns}$, $-20\text{GHz}<\Omega_{1,2}<20$GHz, detuning $\delta_1=50$GHz, $\delta_2=0$ and $B_\text{ext}=0.15$T (all variables in units of $\hbar$), 800000 training epochs and $n=600$ neurons in two fully-connected layers. Using approximated closed system dynamics (Result for full open system in supplemental materials).  }
	\label{NVCenter}
\end{figure*}

The goal is to achieve coherent control between the states $\ket{-1}$ and $\ket{+1}$ of the triplet ground state manifold $\{\ket{-1},\ket{0},\ket{+1}\}$ via coupling to a manifold of excited states. 
Starting from the ground state $\Psi_0=\ket{-1}$, we would like to achieve a general superposition state of the Bloch sphere spanned by the two states
\begin{equation}\label{NVstate}
\Psi(\theta,\phi)=\cos\left(\frac{\theta}{2}\right)\ket{-1}+\expU{i\phi}\sin\left(\frac{\theta}{2}\right)\ket{+1}\, ,
\end{equation}
where $\theta\in\{0,\pi\}$ and $\phi\in\{0,2\pi\}$. These two degenerate ground states are not coupled directly. The degeneracy can be lifted with an external magnetic field $B_\text{ext}$. To coherently control them, we couple the ground states via two lasers to the excited state manifold $\{A_2,A_1, E_X,E_Y,E_1,E_2\}$, which is separated in energy from the ground state triplet within the optical frequency range. Through dissipative couplings (see Methods), a further metastable state can be occupied, thus the NV center is described by an open ten-level system. We investigate the limit where the protocol time is much faster than the dissipation, thus we can approximately treat the system as an effective closed eight-level system (comparison with full dynamics in the Supplemental Materials).
We apply two driving lasers, with time-dependent strengths chosen by the protocol $\Omega_1(t)$, $\Omega_2(t)$. They have a relative detuning $\delta_1$, $\delta_2$ to the energy difference of states $\ket{-1}$ ($\ket{+1}$) and $\ket{A_2}$ (see Methods). 
This system resembles the well known $\Lambda$ system with 3 levels, however with an additional complicated set of levels that has to be excited and controlled. The excited levels interact non-trivially with one another as well. For these systems, simple or analytic solutions are difficult to find, especially for the non-adiabatic regime considered here. 



The goal is to learn protocols to reach arbitrary superposition states of the two levels parameterized by angles $\theta$ and $\phi$. To evaluate how similar protocols are, we introduce the following measures:
Firstly, to evaluate the local change of protocols, we define the norm of the protocol gradient in respect to the target parameters $\theta$ and $\phi$
\begin{equation}\label{grad}
G(\theta,\phi)=||\partial_\theta \beta(\theta,\phi)||_1+ ||\partial_\phi \beta(\theta,\phi)||_1\, , 
\end{equation}
where $\beta(\theta,\phi)$ is the protocol for a given $\theta$ and $\phi$  and $||.||_1$ the $L_1$ norm. This measures how much the protocol is changing across the Bloch sphere. 
Secondly, to measure how close two different protocols are, we define the protocol distance
\begin{equation}\label{dist}
C(\theta,\theta',\phi,\phi')=||\beta(\theta,\phi)-\beta(\theta',\phi')||_1\, , 
\end{equation}
If two protocols are identical, the measure is zero. Else, the protocol distance is positive. Our protocols $\beta$ are a vector of length $3N_T$ consisting of $N_T$ timesteps with the two driving strengths $\Omega_1^{(k)}$, $\Omega_2^{(k)}$ and length of timestep $\Delta t^{(k)}$. To calculate the correlation measure and gradient, we map the parameters between -0.5 and 0.5. 

\begin{figure*}[htbp]
	\centering
	\subfigimg[width=0.7\textwidth]{}{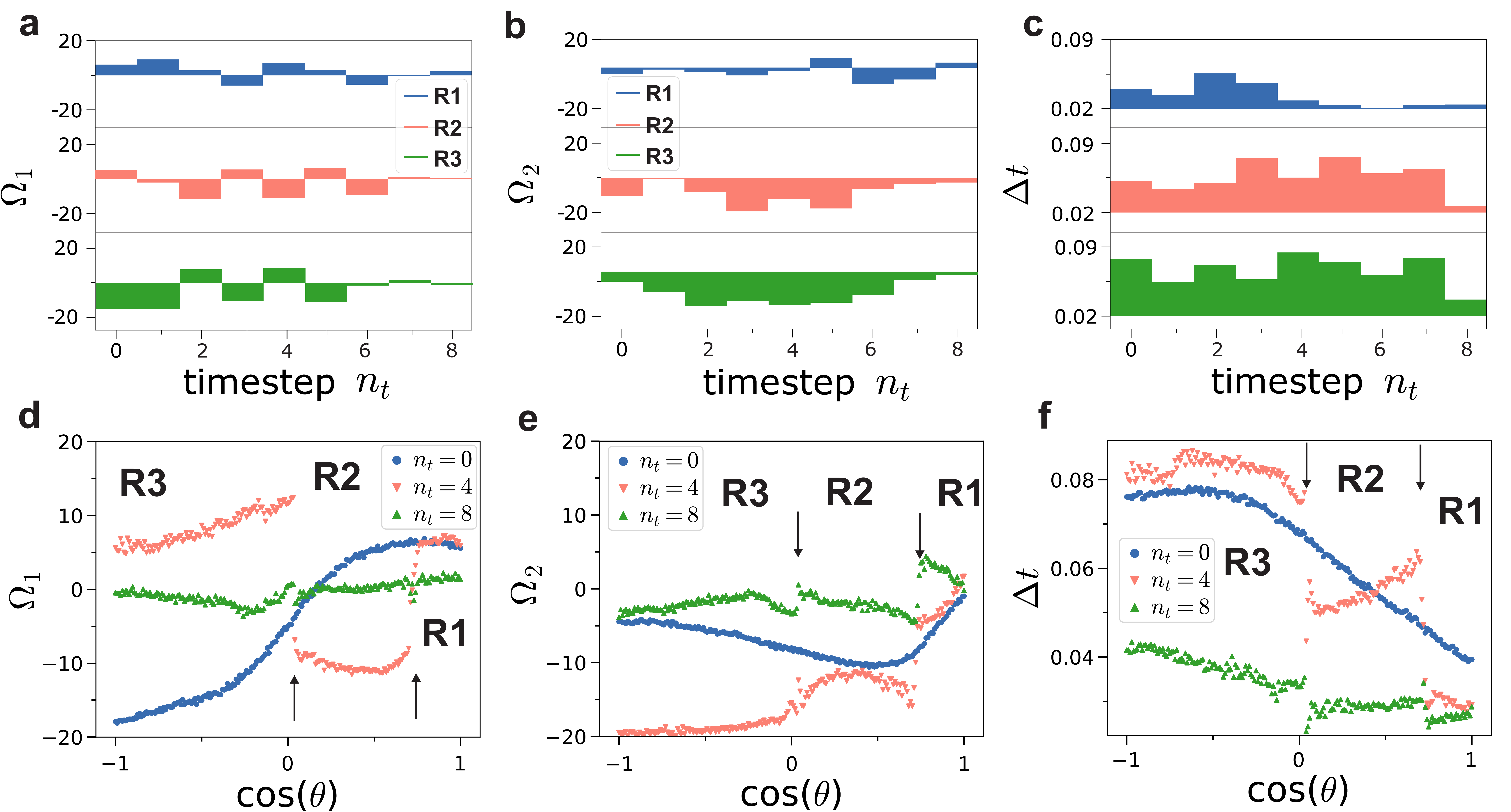}
	\caption{Example optimized driving protocols for target states calculated in Fig.~\ref{NVCenter}. Driving protocols at step $k$ are parameterized by driving strength $\Omega_1^k$, $\Omega_2^k$ and timestep length $\Delta t^k$. \idg{a-c} Example protocols for three representative examples from different protocol `phases' for driving parameter \idg{a} $\Omega_1^k$ \idg{b} $\Omega_2^k$ \idg{c} timestep length $\Delta t^k$. Examples taken from regions as defined in Fig.~\ref{NVCenter}, for R1: $\cos(\theta)=0.9$, $\phi=\pi$, R2: $\cos(\theta)=0.5$, $\phi=\pi$, R3: $\cos(\theta)=-0.5$, $\phi=\pi$. \idg{d-f} Driving protocol for cut along $\theta$ with $\phi=\pi$ for different protocol timesteps $n_t$ and driving parameters \idg{d} $\Omega_1$, \idg{e} $\Omega_2$, \idg{f} $\Delta t$. Protocol show distinct `phases' that jump at $\cos(\theta)\approx0$ and $\cos(\theta)\approx0.75$. See supplemental materials for further example protocols. }
	\label{NVCenterDriving}
\end{figure*}

The result after training is shown in Fig.~\ref{NVCenter}. We see that the fidelity of reaching the target state is high in most areas, with very sharp lines of low fidelity (see Fig.~\ref{NVCenter}a) that divides it into three areas (marked as R1 to R3). The sharpness of these lines increases with number of neurons as the neural network can represent more complex features and abrupt changes (Further demonstration of this feature on a simpler spin rotation problem in Supplemental Materials).  
At these lines, the protocol changes drastically, while it changes only slowly in the other areas. 
We visualize this with the protocol gradient Eq.\ref{grad} in Fig.~\ref{NVCenter}b. It measures how much the protocol changes with $\theta$ and $\phi$. The gradient is low within those three distinct areas, where the protocol is changing only slowly and protocols are very similar. Protocols in these areas are very similar to each other.  We observe specific sharp lines where the gradient is very large. These correlate with the lines of low fidelity in Fig.~\ref{NVCenter}a. Here, the protocol changes drastically, in order to interpolate between one type of protocol to another.
As an analogy, we describe this phenomena as a kind of phase diagram: The areas of low gradient represent `phases' (class of protocols of specific type), that are separated lines of high gradient akin to a `phase transition'. At these lines, the algorithm struggles to find good protocols. 
These `phases' are very well visible in the protocol time $T$, shown in Fig.~\ref{NVCenter}c with distinct areas of similar $T$. These directly reveal some physical insight into the state generation. For states with $\cos(\theta)\approx 1$ (close to initial state), fast protocols are sufficient. For increasing $\theta$, protocols require a longer duration. Our algorithm automatically groups protocols into areas of similar protocol time $T$. Intermediate timed protocols (area R2) is concentrated for $\phi<\pi$, while protocols with long duration (area R3) is mostly located in $\phi>\pi$. This asymmetry comes from the magnetic field $B$ that lifts the degeneracy between the states $\ket{\pm1}$. We also note that $\phi=0$ and $\phi=2\pi$ have different results in terms of fidelity and protocol, although they represent the same quantum state. The reason is that the neural network receives as input only  the angles $\theta$ and $\phi$ of the target state and does not know about the symmetry of the problem (e.g. that $\phi$ is periodic). This symmetry can be enforced by hand, yielding similar fidelities (see Supplemental Materials).  
In Fig.~\ref{NVCenter}d, we show the distances between different protocols using Eq. \ref{dist}. We chose a $21 \times 21$ sampling grid on the Bloch sphere. Although the protocols are actually parameterized by a $3N_\text{T}$ dimensional vector, using the t-SNE algorithm we can map the distances  between different protocols (given by Eq. \ref{dist}) onto a 2D representation. The color indicates the protocol time $T$. We see again distinct clusters of similar protocols, corresponding to the areas marked in Fig.~\ref{NVCenter}b.

Representative examples of driving protocols from the three different protocol `phases' are shown in Fig.~\ref{NVCenterDriving}a-c. The protocols belonging to different `phases' look distinct, while protocols within the same `phase' look very similar (see Supplemental Materials for further example protocols). Fig.~\ref{NVCenterDriving}c shows the length of each time step, where the total time (given by the integral over the steps) increases from R1 to R3. We observe also distinct shapes in the sequence of driving strengths, which vary for R1 to R3 (see Fig.~\ref{NVCenterDriving}a,b). 
In Fig.~\ref{NVCenterDriving}d-f we show the change of the protocol parameters for a cut at ${\phi = \pi}$ along the $\theta$ axis. We again observe the three distinct `phases' for varying $\theta$ in the driving parameters. They change contentiously with $\theta$, however remain similar within one `phase'. However, sudden jumps where the protocol changes drastically are observed at $\cos(\theta)\approx0$ and $\cos(\theta)\approx0.75$ (most visible for timestep $n_t=4$ and $n_t=8$). This indicates the transition to another protocol `phase'.
The protocols found are near-optimal solutions within a complicated optimization landscape~\cite{bukov2018reinforcement}. As such, there are many possible solutions of nearly the same fidelity. By varying hyperparameters of the algorithm or external parameters one may find different types of the protocols in the two-dimensional subspace. However, in most cases there are either two or three distinct areas (see other example in supplemental Materials).

\begin{figure}[htbp]
	\centering
	\subfigimg[width=0.24\textwidth]{a}{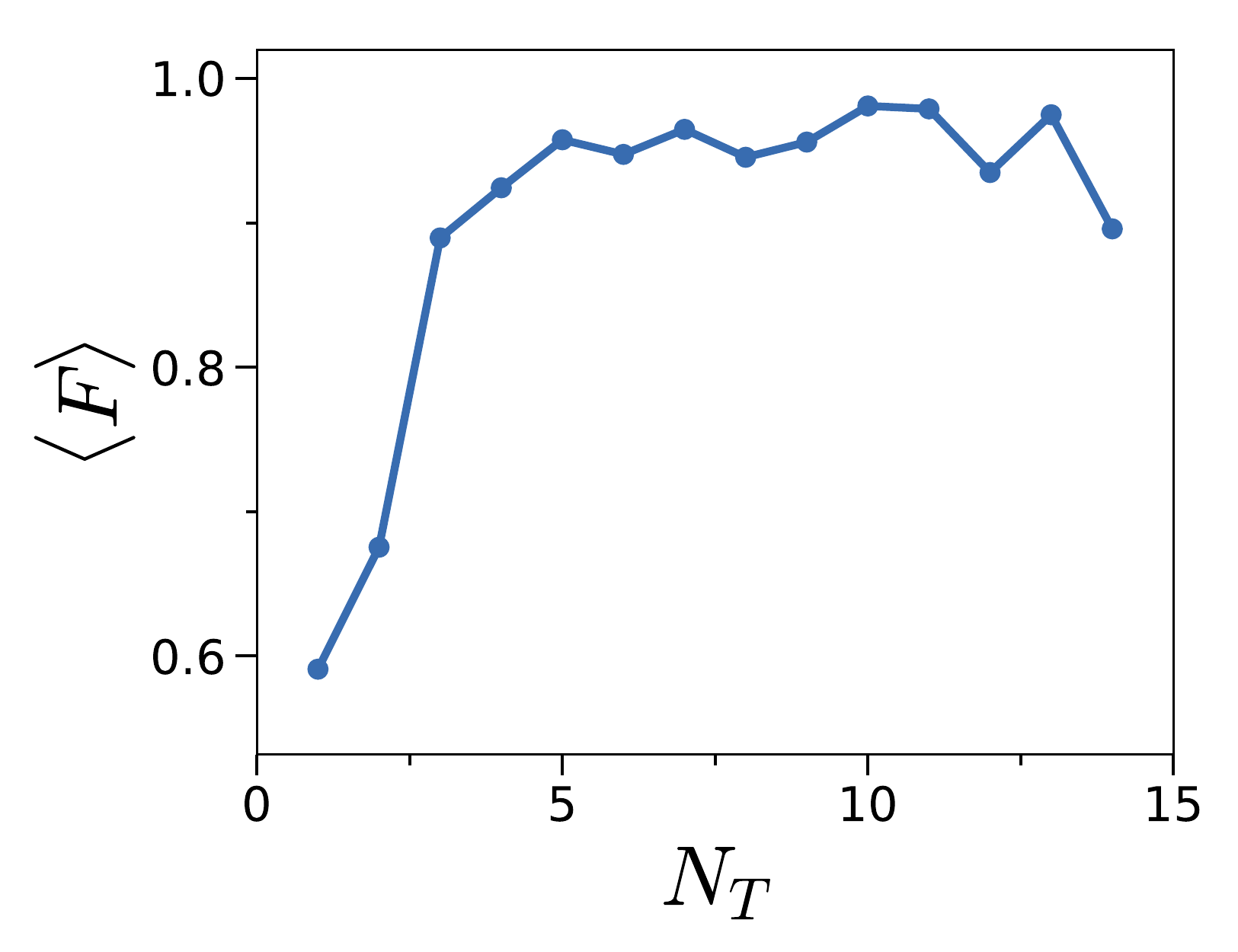}\hfill
	\subfigimg[width=0.24\textwidth]{b}{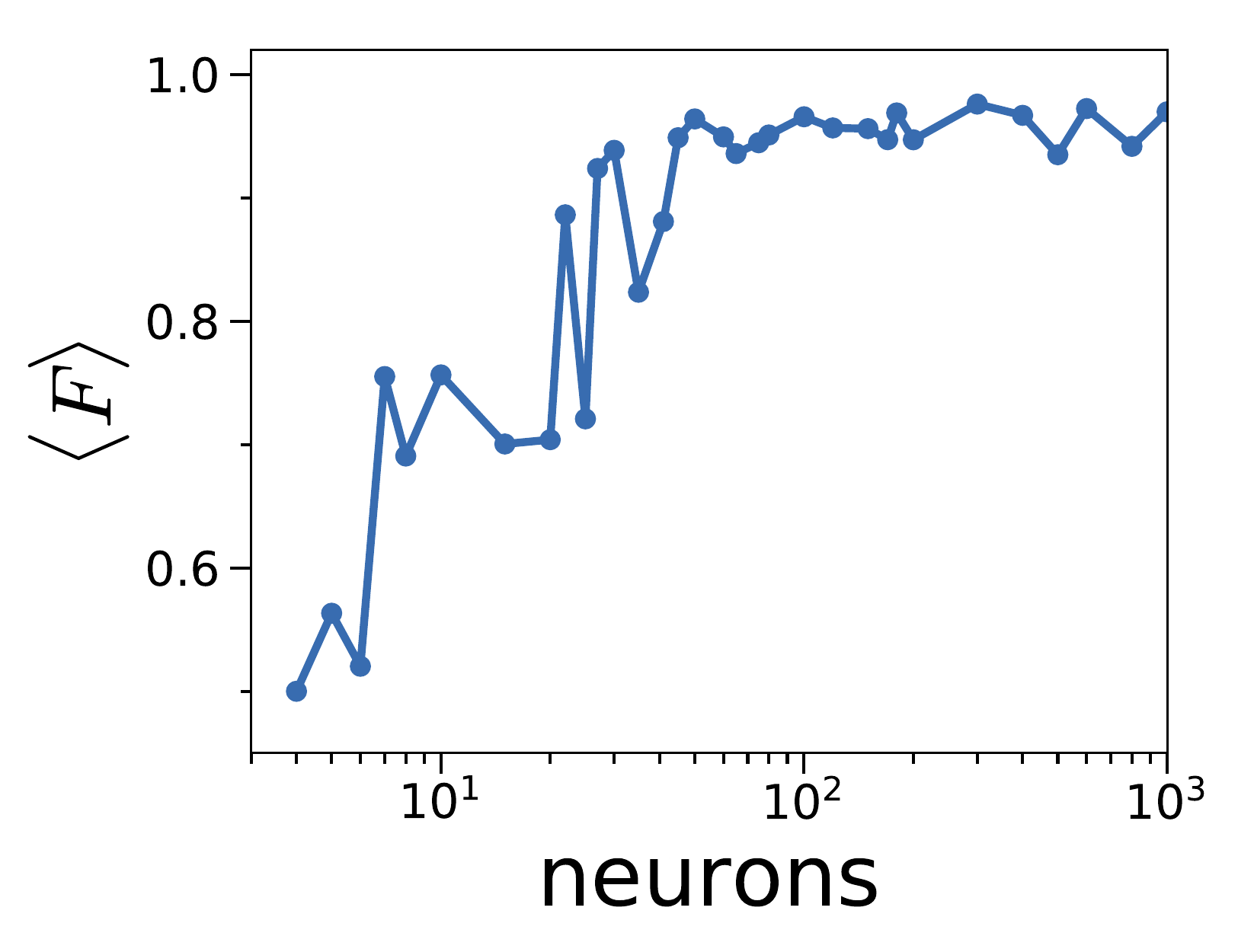}
	\caption{Fidelity by reinforcement learning trained protocols to generate state $\Psi(\theta,\phi)$ averaged over all $\theta$,$\phi$.  \idg{a} Average fidelity $\langle F \rangle$ against number of timesteps for $n=150$ neurons \idg{b} Average fidelity against number of neurons for $N_T=9$ timesteps. Same parameters as in Fig.~\ref{NVCenter}.}
	\label{NVparams}
\end{figure}

The hyperparameters of the physical problem and the reinforcement learning affect the quality of the learning protocols. In Fig.~\ref{NVparams}, we show the average fidelity over all target states for varying number of protocol timesteps (Fig.~\ref{NVparams}a) and number of neural network neurons (Fig.~\ref{NVparams}b). We observe that beyond a certain number of neurons or timesteps, the average fidelity does not increase anymore.

\begin{figure*}[htbp]
	\centering
	\subfigimg[width=0.8\textwidth]{}{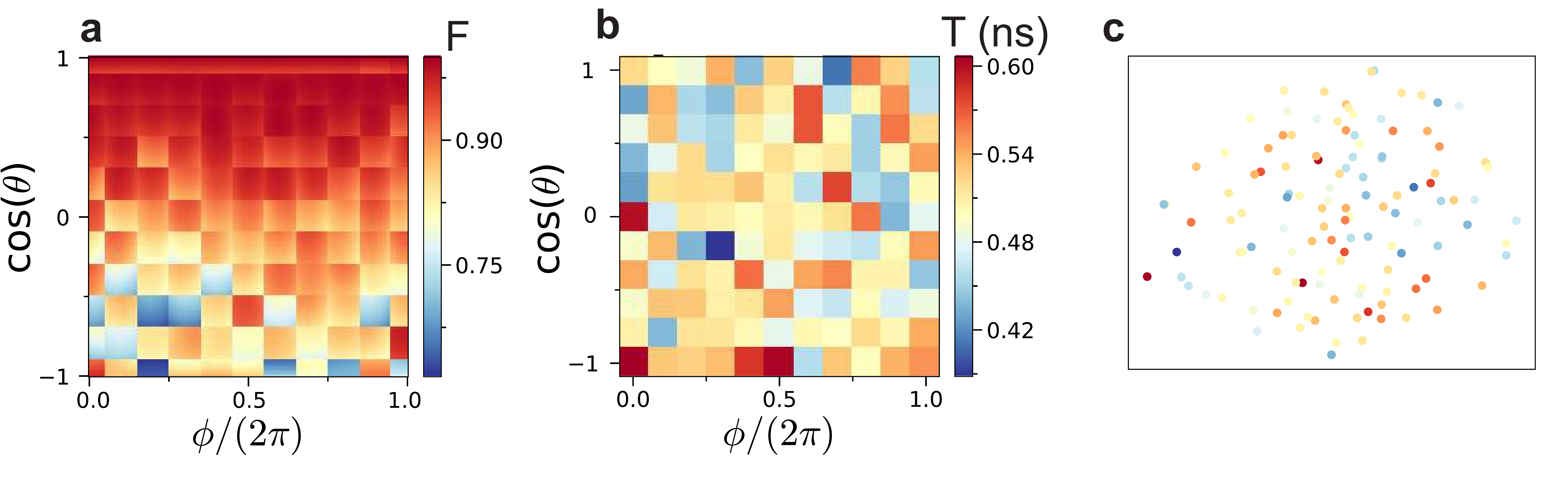}
	\caption{Create arbitrary quantum state in two-level subspace Eq.\ref{NVstate} by optimizing with a standard optimization tool (Nelder-Mead). \idg{a} Fidelity $F$ for preparing state $\Psi(\theta,\phi)$. Optimizing a 11x11 grid over the parameter space, intermediate points interpolated by using nearest protocol (Mean fidelity $\langle F \rangle=0.90$) \idg{b} protocol time $T$. No clustering of protocols is observed.
		\idg{c} Two-dimensional representation of the distance between protocols (determined by Eq.~\ref{dist}) using t-SNE algorithm. Color indicates protocol time of a specific datapoint.
		Same parameters as in Fig.~\ref{NVCenter}. Optimization took in total $5.7\cdot 10^6$ training episodes. Each grid point is optimized with Nelder-Mead for up to 20000 optimization steps, repeated over up to 5 optimization runs starting from a random initially guessed protocol. Optimization is stopped early when fidelity $F>0.99$ is reached.}
	\label{NVCenterNM}
\end{figure*}

To compare with a standard method, we optimize the state preparation with Nelder-Mead (using Python \textit{scipy} library implementation). The result is shown in Fig.~\ref{NVCenterNM}. The two-dimensional space of target states is subdivided into a discrete grid and optimized independently. The achieved average fidelity over all target states is lower compared to the reinforcement learning as it tends to get stuck in local optimization minimas. To reduce this problem, each trained target state was optimized up to 5 times, using random initial guesses. This optimization required nearly an order of magnitude more training episodes compared to the reinforcement learning approach as every grid point is learned independently and sequentially, instead of training all target states at the same time. 
The resulting optimized protocols across $\phi$ and $\theta$ is quite different compared to the one achieved by reinforcement learning. As example, we investigate the protocol time $T$ of the optimized protocols in Fig.~\ref{NVCenterNM}b. For reinforcement learning we find a connected landscape with well defined classes. For the grid optimization, however, there is a large spread in the protocol parameter and $T$ varies largely between different target states, even for neighboring $\theta$ and $\phi$. Thus, a small change in the target state can translate into a very different protocol and  therefore interpolation of protocols between the grid points would yield a very poor state preparation. A two-dimensional representation of the distance between the different protocols reveals no clustering or order (Fig.~\ref{NVCenterNM}c), in contrast to reinforcement learning (Fig.~\ref{NVCenter}c). From this result, we gather the following explanation: The control problem has many nearly equivalent solutions \cite{rabitz2004quantum,bukov2018reinforcement}, which yield similar fidelity but can have widely different control schemes. Reinforcement learning tends to find solutions belonging to the same control class, which produces the observed clusters, whereas Nelder-Mead converges to one of the solutions at random, giving uncorrelated protocol schemes.
It is also worth mentioning that this optimization method takes more training epochs compared to the reinforcement learning method since each grid point is optimized individually. Scaling to a higher density of discretization points or more than two target parameters would require even more training time for the grid approach. The reinforcement learning approach in \cite{niu2019universal} also suffers from similar scalability issues, whereas our deep learning algorithm can overcome such scalability problems as all target states can be trained in parallel.

\section{Discussion}
We demonstrated how to learn all control protocols for a two-dimensional set of quantum states. The neural network is trained using randomly sampled target states. After training, the neural network knows how to generate all possible target states and classifies the near-optimal protocols automatically in specific groups, e.g. the time needed to generate specific states as well as the driving strengths. The clustering feature could be useful tool to identify physical principles and find generalized driving protocols. For practical usage, the clustering of solutions is also advantageous as the driving protocols within a cluster do not need to change drastically for a small change in target state.
The drop in fidelity at the boundary between two clusters is a result of the finite capacity of neural networks to represent sharp changes and could be mitigated by choosing a discontinuous activation function. 
However, the exact mechanism how the clusters are found by the neural network remains unclear and requires further studies.
To speed up training, our algorithm could be easily parallelized by calculating multiple target states at the same time. 
This is more difficult for other optimization techniques such as transfer learning as it requires input from previously trained neural networks and does not scale well with number of sampling points as well as the dimension of the problem (ie. number of target parameters) \cite{niu2019universal}. 

Using our algorithm, we showed how to optically control the electron spin in a complex multi-level NV center. The electron spin is only indirectly coupled via several other interacting levels, making it difficult to construct good control pulses. Our method gives us access to a tailored protocol for any superposition state. 
With our method, we can create arbitrary states within a short time $T\approx 0.5 \text{ns}$, avoiding slower dissipative processes and superseding other adiabatic-type protocols to create superposition states \cite{zhou2017accelerated}. Our protocol requires only 9 timesteps to achieve high fidelity for arbitrary states, which is considerably less than comparable approaches \cite{tian2019optimal,porotti2019coherent}. Our results indicate that even less timesteps could be sufficient to reach high fidelity (see Fig.\ref{NVparams}a). 
Our results could help in achieving quantum processor based on NV centers \cite{chen2020optimisation} and could be applied to various other quantum processing platforms and optimal control \cite{werschnik2007quantum}.

Our neural network based algorithm learns the full two-dimensional set of target states. The same concept could help to identify control unitaries for continuous variable quantum computation \cite{hillmann2020universal} or serve as an alternative to transfer learning in quantum neural network states \cite{zen2019transfer}. 
The neural network finds patterns, which may be useful to identify phase transitions in quantum control \cite{bukov2018reinforcement}, as well as identify physical concepts in the way the protocols create quantum states \cite{nautrup2020operationally,iten2020discovering}. The resulting classification of protocols also opens up the potential of using reinforcement learning in identifying phase transitions in physical systems \cite{PhysRevB.94.195105,Rem2019,Ming2019}.
Furthermore, our approach could help correcting drifts in superconducting qubits \cite{2020arXiv200108343F}. Normally, each target unitary has to be retrained individually which may take a lot of time for many gates. Our method allows one to retrain all the protocols at the same time, which may save time to correct errors and drifts.

\section{Methods}
\subsection{NV center}
\begin{figure}[htbp]
	\centering
	\subfigimg[width=0.3\textwidth]{}{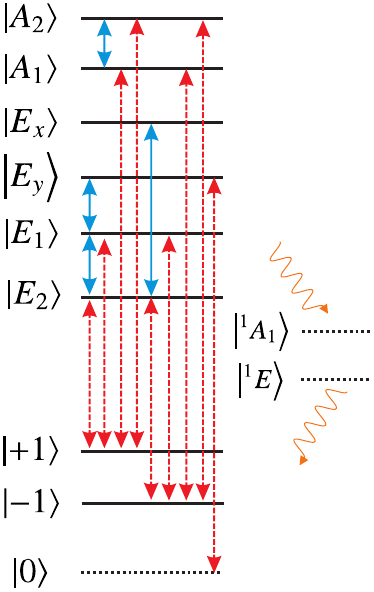}
	\caption{Level structure of the NV center. Consists of two target ground state levels $\{\ket{-1}, \ket{+1}\}$, third ground state level $\ket{0}$, excited states $\{\ket{A_2}, \ket{A_1}, \ket{E_X}, \ket{E_Y}, \ket{E_1}, \ket{E_2}\}$ and metastable states $\{\ket{^1 A_1},\ket{^1 E}\}$. Laser coupling of levels is shown as red dashed lines, coupling via internal level structure as blue solid lines. For short driving protocols $\{\ket{^1 A_1},\ket{^1 E}\}$ and $\ket{0}$ are disregarded as approximation.}
	\label{NVCenterLevel}
\end{figure}

In this paper we consider the ten-level model of the NV$^{-}$ center, which comprises three ground states, six excited states and one metastable state. The ground states form a spin-1 triplet with a zero-field splitting of $D_{gs} \approx 2 \pi \times 2.88$ GHz between the $m_s = \pm 1$ and $m_s = 0$ sublevels. The energy gap of $E_g = 1.94$ eV between the ground states and excited states due to Coulomb interaction gives rise to the well-known zero-phonon line (ZPL) optical transition. The level structure is shown in Fig.\ref{NVCenterLevel}.

The ground state Hamiltonian can be written in the basis $\{ \ket{-1}, \ket{0}, \ket{+1} \} \equiv \{\ket{m_s = -1}, \ket{m_s = 0}, \ket{m_s = +1} \}$ as (setting $\hbar = 1$ and the energy of $\ket{0}$ as zero)~\cite{tian2019optimal}
\begin{equation}
H_\text{gs} = (D_\text{gs} - g_\text{gs} \mu_\text{B} B_{\text{ext}} ) \ket{-1} \bra{-1} + (D_{gs} + g_{gs} \mu_\text{B} B_{\text{ext}} ) \ket{+1} \bra{+1} 
\end{equation}
where $g_\text{gs} = 2.01$ is the Land\'e $g$-factor for the ground state, $\mu_\text{B}$ is the Bohr magneton, and $B_\text{ext}$ is the external magnetic field applied along the NV quantization axis. The magnetic field splits the degeneracy of the states $\ket{-1}$ and $\ket{+1}$. Considering low temperatures, the phononic effects in the diamond are suppressed (and not considered here), while the splittings in the excited state due to spin-spin and spin-orbit interactions become significant. Taking into account these interactions, the excited-state Hamiltonian can be written in the basis of $\{ \ket{A_2}, \ket{A_1}, \ket{E_X}, \ket{E_Y}, \ket{E_1}, \ket{E_2} \}$ as
\begin{equation}
H_\text{es} = E_g I + \left(\begin{matrix}H_1&0\\0&H_2\end{matrix}\right)
\end{equation}
where $I$ is the $6 \times 6$ identity matrix and
\begin{equation}
H_1 = \left(\begin{matrix}\Delta+2l_z&g_\text{es}\mu_\text{B} B_\text{ext}\\g_\text{es}\mu_\text{B} B_\text{ext}&-\Delta+2 l_z\end{matrix}\right)
\end{equation}
\begin{equation}
H_2 = \left(\begin{matrix}-D_\text{es}+l_z&0&0&\Delta^{\prime\prime}\\0&-D_\text{es}+l_z&i\Delta^{\prime\prime}&0\\0&-i\Delta^{\prime\prime}&0&-g_\text{es}\mu_\text{B} B_\text{ext}\\\Delta^{\prime\prime}&0&-g_\text{es}\mu_\text{B} B_\text{ext}&0\end{matrix}\right)
\end{equation}
describe the level splittings in the excited-state manifold with $\Delta = 2\pi \times 1.55$ GHz, $D_\text{es} = 2\pi \times 1.42$ GHz and $\Delta^{\prime\prime} \approx 2\pi \times 0.2$ GHz denotes the spin-spin interactions. $l_z = 2\pi \times 5.3$ GHz is the axial spin-orbit splitting, and $g_{es} \approx 2.01$ is the Land\'e $g$-factor for the excited state.

Coherent control of the NV$^{-}$ center is accomplished by applying two laser fields with frequencies $\omega_1$ and $\omega_2$ and Rabi frequencies $\Omega_1$ and $\Omega_2$. 

The electric-dipole coupling between the ground and excited states is given by the interaction Hamiltonian
\begin{equation}
H_\text{int} = \left(\begin{matrix}0&v\\v^\dag&0\end{matrix}\right)
\end{equation}
where 
\begin{equation}
v = \left(\begin{matrix}i\epsilon_x&-i\epsilon_x&0&0&-i\epsilon_x&-i\epsilon_x\\0&0&0&2\epsilon_x&0&0\\-i\epsilon_x&-i\epsilon_x&0&0&i\epsilon_x&-i\epsilon_x\end{matrix}\right)
\end{equation}
is the $3 \times 6$ coupling matrix with the rows forming the ground state basis and the columns forming the excited state basis. $\epsilon_x = 2\Omega_1 \cos (\omega_1 t) + 2\Omega_2 \cos (\omega_2 t)$. 
The total Hamiltonian including driving is given as $H_\text{NV}=H_\text{gs}+H_\text{es}+H_\text{int}$.
We now move into the rotating frame by transforming the Hamiltonian to the interaction picture $H_I=\expU{iH_\text{Eg}t}\left(H_\text{tot}-H_\text{Eg}\right)\expU{-iH_\text{Eg}t}$, with $H_\text{tot}=H_\text{gs}+H_\text{es}+H_\text{int}$ and $H_\text{Eg}=E_\text{g}\sum_{k=4}^9\ket{k}\bra{k}$. Neglecting the counter-rotating terms, the $\epsilon_x$ terms in the interaction Hamiltonian are replaced by
$\epsilon_x' = \Omega_1 \cos (\delta_1 t) + \Omega_2 \cos (\delta_2 t)$, with the detuning $\delta_i = \omega_i-E_\text{g}$, $i = 1,2$.
In addition, there is a metastable state $\ket{m}$ in the Hamiltonian, totaling to a 10 level system.
The NV center is subject to dissipation via decay of excited states as shown in Table \ref{Diss}. 
The full system including dissipation is solved using the Lindblad Master equation \cite{breuer2002theory}
\begin{equation}\label{LindbladEq}
\pdif{\rho}{t}=-\frac{i}{\hbar}\left[H,\rho\right]-\frac{1}{2}\sum_m\left\{\cn{L}{m}\an{L}{m},\rho\right\}+\sum_m\an{L}{m}\rho\cn{L}{m}~,
\end{equation}
where $\an{L}{m}$ are the Lindblad operators describing the dissipation. 
\begin{table}[htbp]
	\centering
	\begin{center}
		\begin{tabular}{|c|c|}
			\hline
			\textbf{Transition}                                     & \textbf{Decay rate (ns$^{-1}$)} \\ \hline
			$\ket{A_2}, \ket{A_1},\ket{E_1},\ket{E_2} \to \ket{+1}$ & 1/24                            \\ \hline
			$\ket{A_2}, \ket{A_1},\ket{E_1},\ket{E_2} \to \ket{-1}$ & 1/31                            \\ \hline
			$\ket{A_2}, \ket{A_1},\ket{E_1},\ket{E_2} \to \ket{0}$  & 1/104                           \\ \hline
			$\ket{A_2}, \ket{A_1},\ket{E_1},\ket{E_2} \to \ket{m}$  & 1/33                            \\ \hline
			$\ket{E_x}, \ket{E_y} \to \ket{0}$                      & 1/13                          \\ \hline
			$\ket{E_x}, \ket{E_y} \to \ket{+1}, \ket{-1}$           & 1/666                           \\ \hline
			$\ket{E_x}, \ket{E_y} \to \ket{m}$                      & 0                              \\ \hline
			$\ket{m} \to \ket{0}$                                   & 1/303                           \\ \hline
			$\ket{m} \to \ket{+1}, \ket{-1}$                        & 0                               \\ \hline
		\end{tabular}
		\caption{Decay channels and rates for the NV center. $\ket{m}$ represents the metastable state which comprises the singlet states $\ket{^1 A_1}$ and $\ket{^1 E}$.}\label{Diss}
	\end{center}
\end{table}

For timescales shorter than the fastest dissipation channel, the effect of dissipation can be neglected. For the NV center, this would be for $T\ll13\text{ns}$. In this limit, the dynamics is effectively governed by the coherent Hamiltonian only and reduces to a 8 level system, as both state $\ket{0}$ and  $\ket{m}$ are only accessible via dissipation.

\subsection{Deep reinforcement learning}
Here, we describe our machine learning algorithm in more detail. 
We learn the driving protocol via a deep Q-learning network~\cite{mnih2015human}, utilizing the actor-critic method with Proximal policy optimization (PPO) \cite{schulman2017proximal}. We use \cite{spinningup} implementation of the algorithm in Tensorflow~\cite{tensorflow2015-whitepaper}.
The quantum system is controlled by an agent, that depending on the state $s_t$ of the system acts with an action $a_t$ (e.g. driving parameters for time $t$) using the probabilistic policy $\pi(a_t|s_t)$. At every timestep a reward (e.g. the fidelity of quantum state) is paid out. The goal is to repeatedly interact with the quantum system and learn the best policy that gives the highest final reward. One normally starts with a random policy, that explores many possible trajectories. Over the course of the training, the policy is refined and converges (hopefully) to the optimal (deterministic) policy. 
However, optimizing the policy directly can be difficult, as one round of the protocol is played out over $N_T$ timesteps. Here, the question is how to optimize the policy at each step such that one finds the optimal final reward and not become stuck in local maximas?

Here, it has been shown one can overcome the difficulties of policy learning with Q-learning. The idea is to find the Q-function $Q_\pi(s_t,a_t)$ that estimates the future reward (from the point of timestep $t$) that is paid out at the end the full protocol with this policy. The goal is to learn a policy that prioritizes long-term rewards over smaller short-term gains. 
The optimal Q-function is determined by the Bellman equation
\begin{align*}
Q(s_t,a_t,\pi)&=\mathbb{E}\left[r_t+\gamma Q(s_{t+1},a_{t+1},\pi)\right]\\
&=\mathbb{E}\left[r_t+\gamma r_{t+1}+\gamma^2 r_{t+2}+\dots\right]\,
\end{align*}
where $\mathbb{E}[.]$ indicates sampling over many instances. $\gamma\le1$ is a discount factor that weighs  future rewards against immediate rewards.

Proximal policy optimization is based on the idea of combining both policy and value learning in the actor-critic method.
The idea is to have two neural networks: a policy network and a value network. The policy network (actor) decides on the next action by determining the parameters of the policy. The value based network (critic) evaluates the taken action on how well it solves the task and estimates the future expected reward. It is used as an input to train the policy network. 

Better performance can be achieved if the Q-function is split into two parts~\cite{wang2015dueling}: $Q(s_t,a_t)=A(s_t,a_t)+V(s_t)$, where $A(s_t,a_t)$ is the advantage function and $V(s_t)$ the value function. $V(s_t)$ gives the expected future reward averaged over the possible actions according to the policy. This is the output of the critic network. $A(s_t,a_t)$ gives the improvement in reward for action $a_t$ compared to the mean of all choices.

Learning is achieved by optimizing the network parameters with a loss function via gradient descent~\cite{kingma2014adam}.
The loss function of the value network is the square of the difference of the value function of the network and the predicted reward in the next timestep $L_\text{V}(\theta)=\mathbb{E}_t\left[(V_\theta(s_t)-y_t)^2\right]$, where $\theta$ are the current network parameters, $y_t=r_t+V_{t+1}$, where $V_{t+1}$ is the output of the value network for the next timestep (it is set to zero if this is the last timestep).  

The advantage function $A(s_t,a_t)$ tells us how good a certain action $a_t$ is compared to other possible actions. The advantage function is the input to train the policy network (the actor). Following the idea of proximal policy optimization~\cite{schulman2017proximal}, the goal is to minimize the loss function of the policy network
\begin{equation}
L_\text{p}(\theta)=-\mathbb{E}_t\left[\frac{\pi_\theta(s_t,a_t)}{\pi_{\theta_\text{old}}(s_t,a_t)}A(s_t,a_t)\right]\, ,
\end{equation}
where $\theta$ are the network parameters and $\theta_\text{old}$ are the network parameters of a previous instance. Maximizing $L_\text{p}(\theta)$ for the network parameters $\theta$ over many sampled instances guides the distribution $\pi_\theta(s_t,a_t)$ such that it returns actions $a_t$ with maximal advantage. However, the ratio 
\begin{equation*}
b_t(\theta)=\frac{\pi_\theta(s_t,a_t)}{\pi_{\theta_\text{old}}(s_t,a_t)}
\end{equation*}
can acquire excessive large values, causing too large changes in the policy in every training step and making convergence difficult. For PPO, it was proposed to use a clipped ratio $\epsilon$~\cite{schulman2017proximal}
\begin{equation*}
L_\text{p}(\theta)=-\mathbb{E}_t\left[\text{min}\left\{b_t(\theta)A(s_t,a_t),\text{clip}(r_t(\theta),1-\epsilon,1+\epsilon )A(s_t,a_t)\right\}\right]\, ,
\end{equation*}
such that the update at each step stays in reasonable bounds.

The input state $s_t$ to the neural network are the wavefunction $\Psi(t_n)$ (or density matrix $\rho(t_n)$ at previous timestep $n$, as well as a random target state $\Psi_\text{target}$. The output of the policy network are the parameters for the policy $\pi(a_t|s_t,\mu,\sigma)$, where the actions (pulse amplitude and time step duration) are sampled from a normal distribution with mean value $\mu$ and width $\sigma$. $\mu$ is chosen by the neural network (given the input state) and $\sigma$ is a global variable that is initially  large (ensuring that driving parameters are initially sampled mostly randomly to explore many possible trajectories). It is optimized via the loss function and decreases over the training, until at convergence it is close to zero and the sampled driving parameters converge to $\mu$.  We constrain the possible output values for the driving parameters by punishing values outside the desired range with a negative reward.

We optimize the neural network over many epochs $N_\text{E}$. 
In each epoch,  the Schr\"odinger equation (or master equation) is propagated for a total time $T$ with $N_T$ discrete timesteps of width $\Delta t$, with respective times $t_n$.  For one epoch, the system runs the network $N_T$ times. From the policy network the driving parameters for the ${n+1}$ timestep are sample. The output of the value network is used to train the policy network.
Each network is composed of two hidden layers of fully connected neurons of size $N_\text{H}$ with ReLu activation functions.  
The neural networks are trained with the loss function after calculating the full time evolution to time $T$ over $N_T$ timesteps. Training data is sampled from a buffer storing earlier encountered trajectories.
For the actual implementation, we choose the following parameters: learning rate for both value and policy network $\alpha=10^{-5}$, training over $N_\text{E}=800000$ epochs and clip ration $\epsilon=0.05$. Out of bounds driving parameters are punished by a factor of 0.2.

To improve the mean fidelity over all target states, we choose the target states not completely random, but biased towards areas of lower fidelity. This is achieved by laying a $20 \times 20$ grid across the $\theta$ and $\phi$ space, then binning the last 10000 results and the achieved fidelity. We choose the next target state by sampling the probability distribution 
\begin{equation*}
P(\theta,\phi)=\frac{1}{2\mathcal{N}}(\eta+(1-\eta)(1-\mean{F(\theta,\phi)}))\, ,
\end{equation*}
where $\mathcal{N}$ is a normalization factor and $\eta$ determines how strongly the sampling is biased towards low fidelity. We choose $\eta=0.5$.

\section{Data availability}
Data are available from the authors on reasonable request.

\section{Code availability}
Computer code is accessible online on Github in \cite{haug2020fullcontrol}.

\begin{acknowledgments}
\section{Acknowledgments}
The computational work for this article was partially performed on resources of the National Supercomputing Centre, Singapore (https://www.nscc.sg).
\end{acknowledgments}

\bibliography{library}

\appendix

\section{Control of the NV center}
\begin{figure}[htbp]
	\centering
	\subfigimg[width=0.3\textwidth]{}{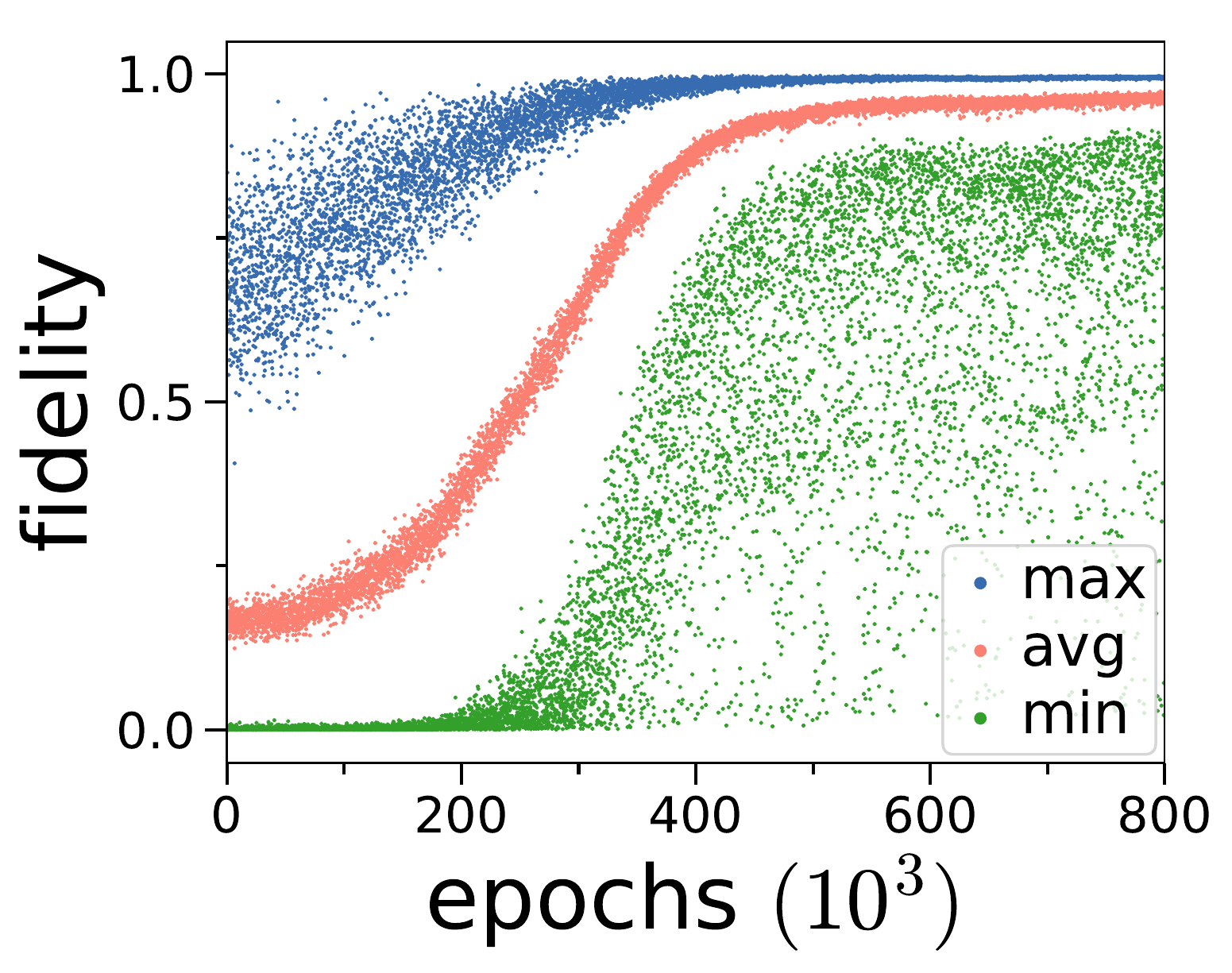}
	\caption{Learning trajectory over 800000 epochs for Fig.~2 of main text. Plotted is the maximal, minimal and average fidelity, calculated over bins of 100 epochs. }
	\label{NVCentLearning}
\end{figure}

Here we present further data on learning control protocols for the NV center. We show the learning trajectory in Fig.~\ref{NVCentLearning}. Initially, the neural network chooses random driving parameters, resulting in low fidelity. After training for some time, the average fidelity increases until it converges. The minimal fidelity observed over all target states remains quite low. This the result of the characteristic lines of low fidelity.

\begin{figure*}[htbp]
	\centering
	\subfigimg[width=0.95\textwidth]{}{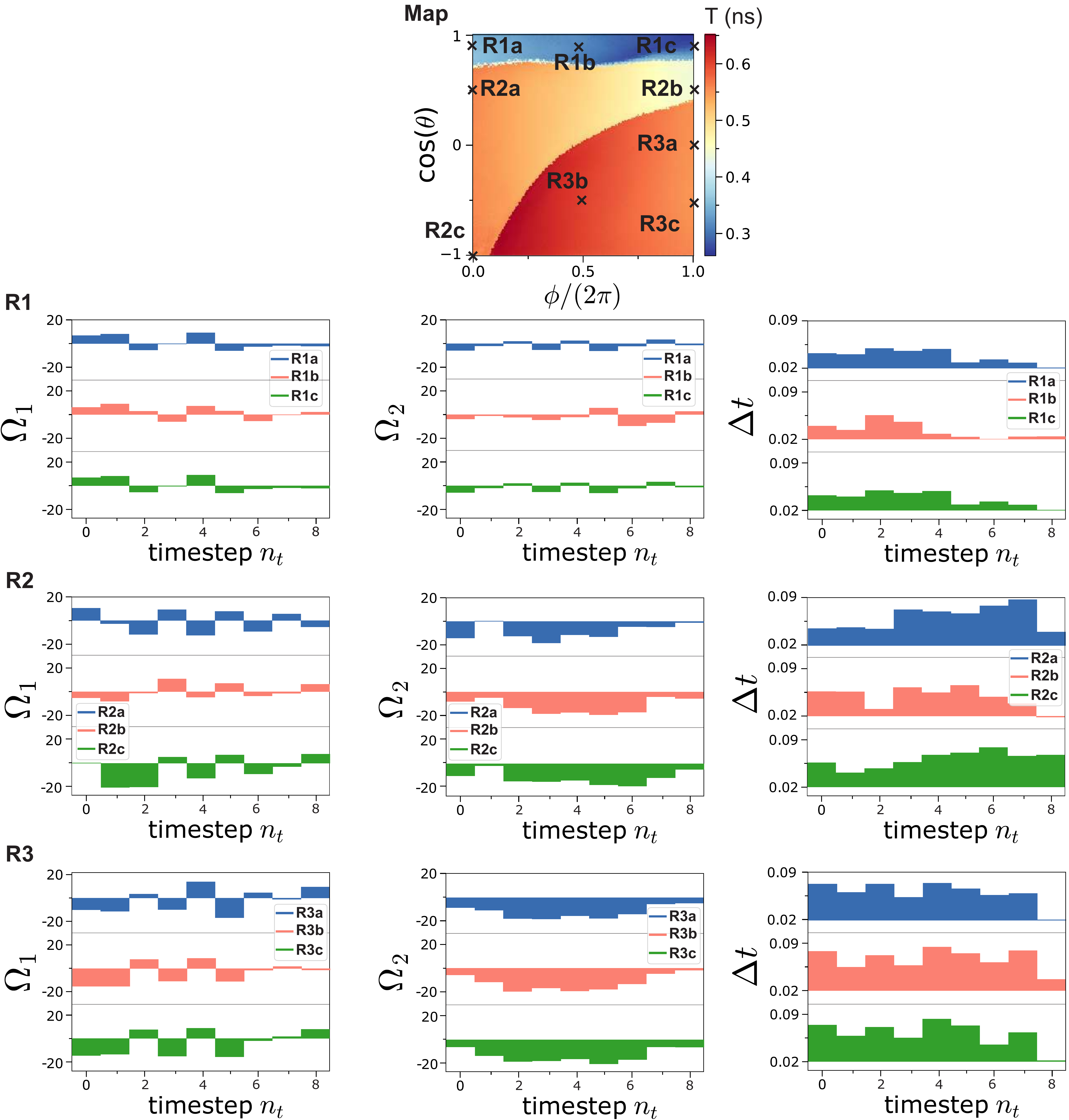}
	\caption{Further example optimized driving protocols for target states calculated as calculated in Fig.~2 of main text. Driving protocols at step $k$ are parameterized by driving strength $\Omega_1^k$, $\Omega_2^k$ and timestep length $\Delta t^k$. 
		The graph at the top shows the location of the various examples within the protocol landscape. The neural network grouped the driving protocols in three distinct areas. We show three examples from the three areas R1, R2 and R3. Note that protocols taken from one specific area are relatively similar (even when far apart in terms of $\theta$, $\phi$), while they are very different compared to protocols from other areas.  }
	\label{NVCentExamples}
\end{figure*}

Fig.~\ref{NVCentExamples} shows further example driving protocols over the protocol landscape. The example protocols are sampled from various points in the three regions R1, R2 and R3. 

\begin{figure}[htbp]
	\centering
	\subfigimg[width=0.24\textwidth]{a}{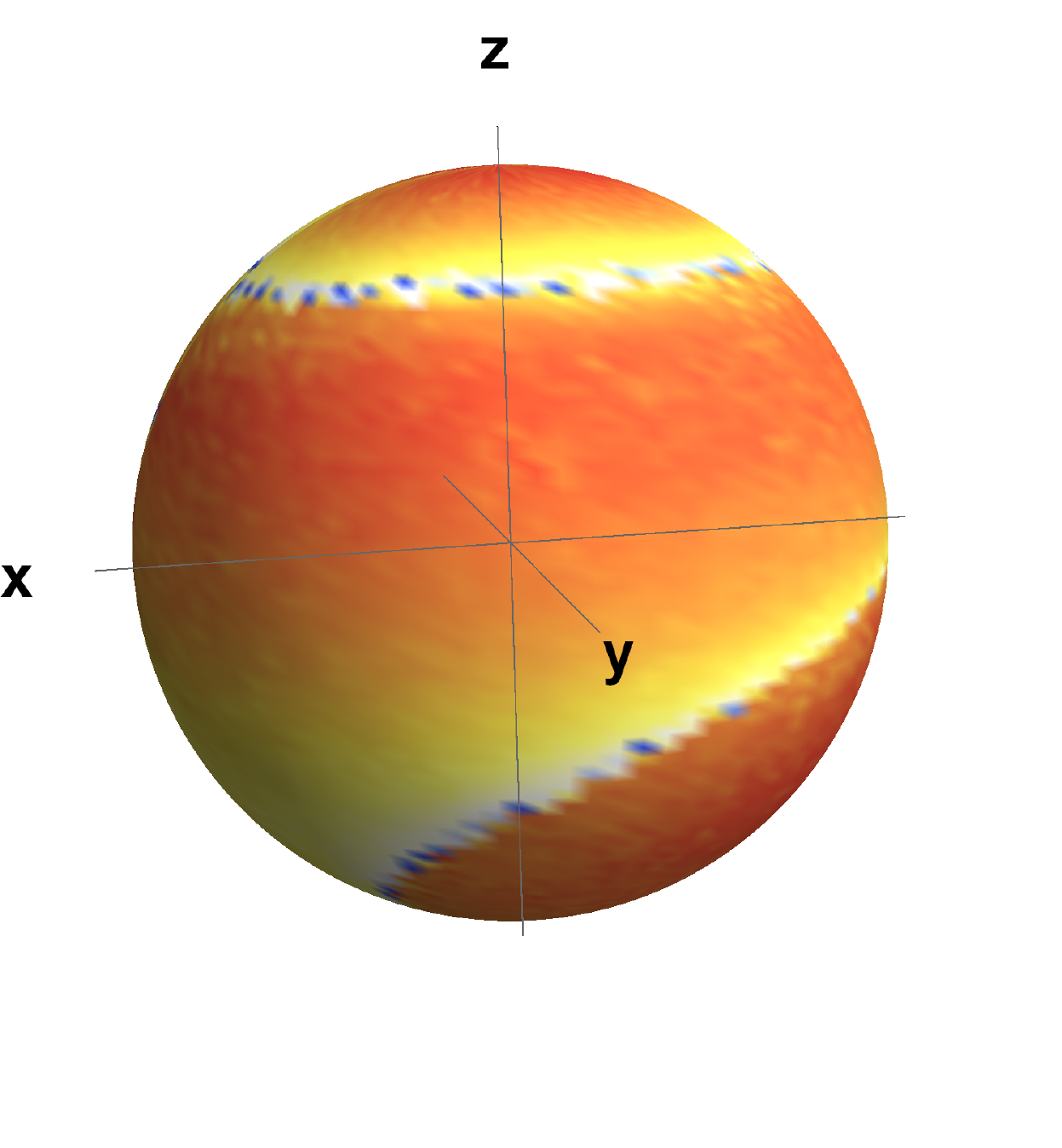}\hfill
	\subfigimg[width=0.24\textwidth]{b}{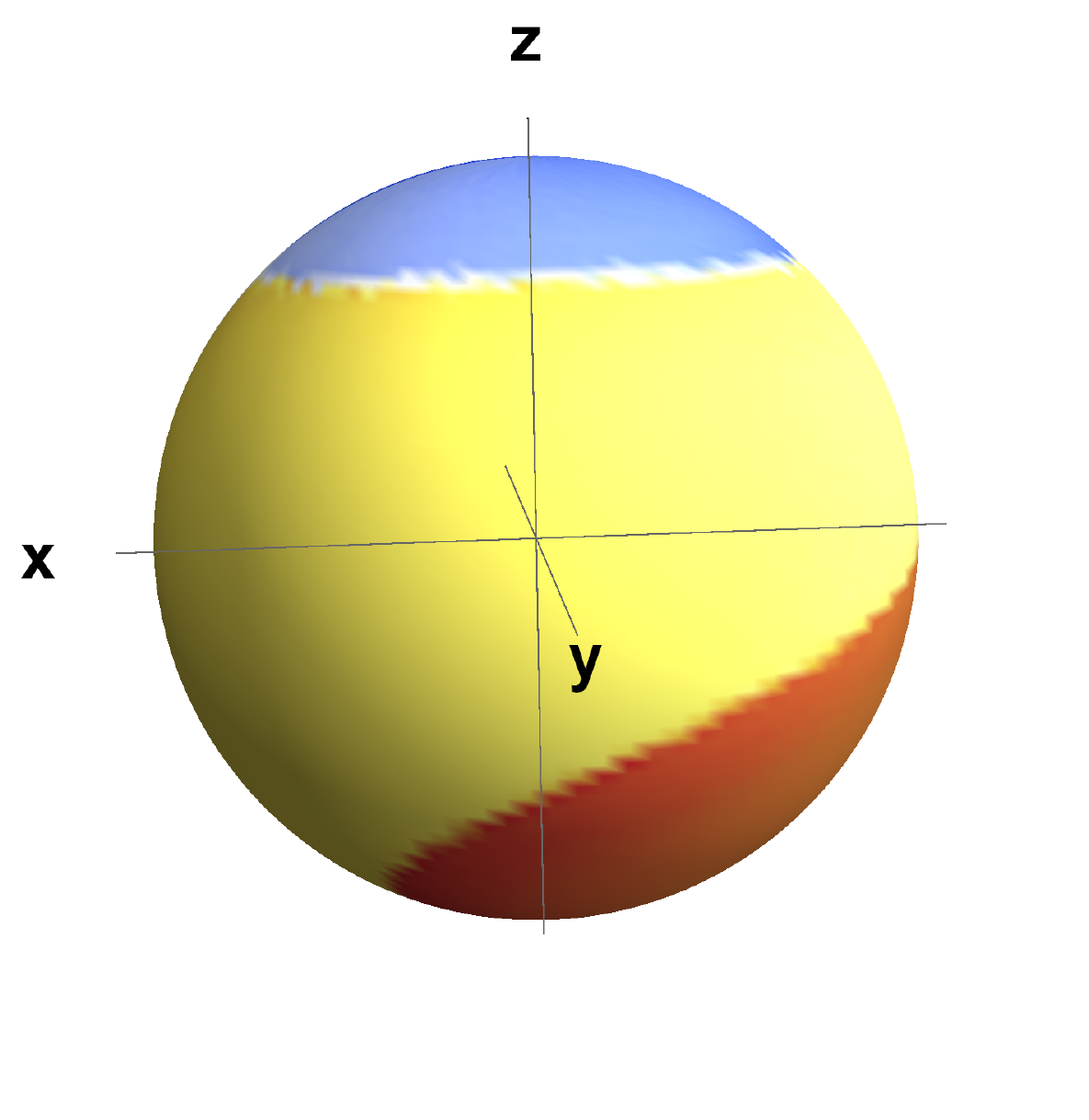}
	\caption{Plot the protocol after training with neural net for target states $\theta$, $\phi$ on the Bloch-sphere. Shows the same results as in Fig.2 of main text. \idg{a} Fidelity $F$ of target state. \idg{b} protocol time $T$. }
	\label{NVBloch}
\end{figure}

Finally, we show the fidelity $F$ and protocol time $T$ of the main text mapped onto the Bloch-sphere. The result is shown in Fig.~\ref{NVBloch}

\section{Fidelity for open system}
\begin{figure}[htbp]
	\centering
	\subfigimg[width=0.24\textwidth]{a}{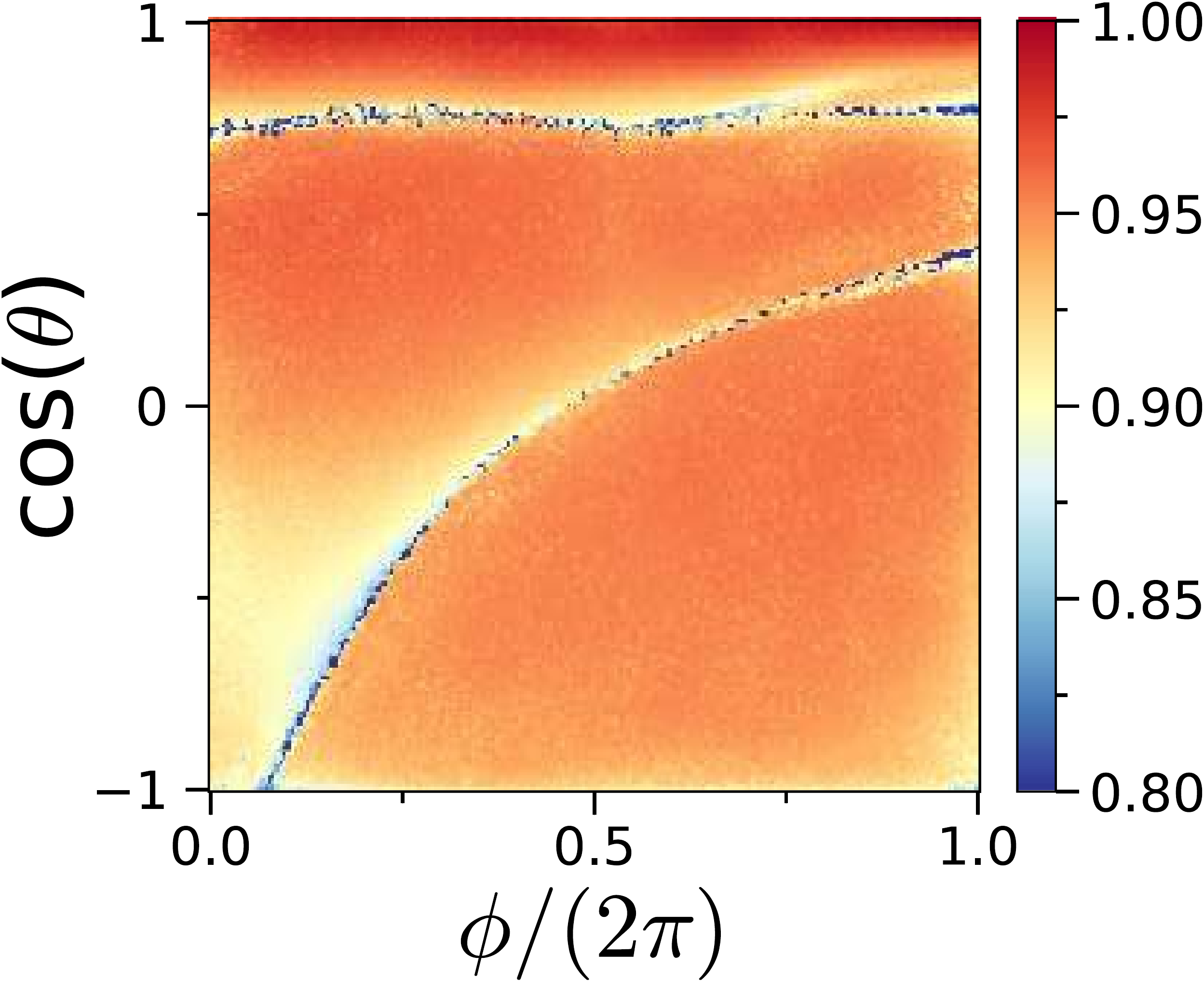}\hfill
	\subfigimg[width=0.24\textwidth]{b}{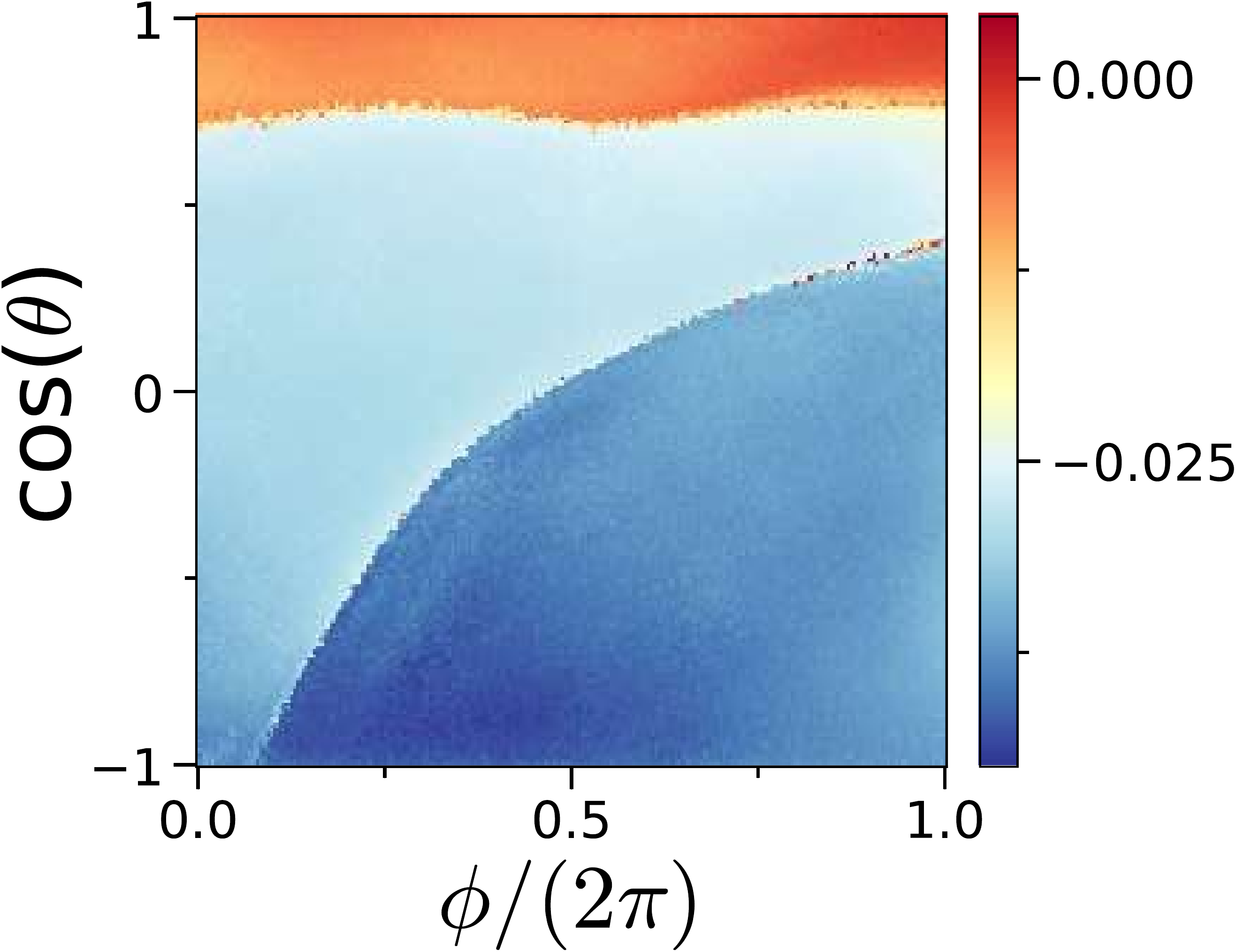}
	\caption{Create arbitrary quantum state in two-level subspace (see Eq.~2 in main text) using the Lindblad equations with dissipative 10 level system and reinforcement learning. Using same protocol as trained for closed quantum system as in main text. \idg{a} Fidelity of preparing state $\Psi(\theta,\phi)$ (Mean fidelity $\langle F \rangle=0.949$) \idg{b} Difference in fidelity between closed quantum system and open quantum system.
		Parameters: $N_\text{T}=9$ timesteps, variable time per step with maximal time $0.2\text{ns}<T<0.8\text{ns}$, $-20<\Omega_{1,2}<20\text{GHz}$, 800000 training epochs, detuning $\delta_1=50\text{GHz}$, $\delta_2=0$ and $B_\text{ext}=0.15T$. }
	\label{NVCenterFull}
\end{figure}

In the main text, we optimize the fidelity of generating superposition states in NV center. We use the approximated closed system dynamics of the NV center. This is a valid approximation for time scales much shorter than the dissipative channel, a condition fulfilled in our case. In Fig.~\ref{NVCenterFull}, we show the result for the full open system quantum dynamics. We use the same protocol as optimized for the closed system. The mean fidelity is slightly reduced for the open system dynamics. We observe that the fidelity for each protocol class is affected differently by the open system dynamics (see Fig.\ref{NVCenterFull}b). The first class actually improves, whereas the other classes have decreased fidelity. It directly correlates with the protocol time $T$, meaning that a longer protocol is more affected by the dissipation.

\section{Magnetic field}
\begin{figure}[htbp]
	\centering
	\subfigimg[width=0.24\textwidth]{a}{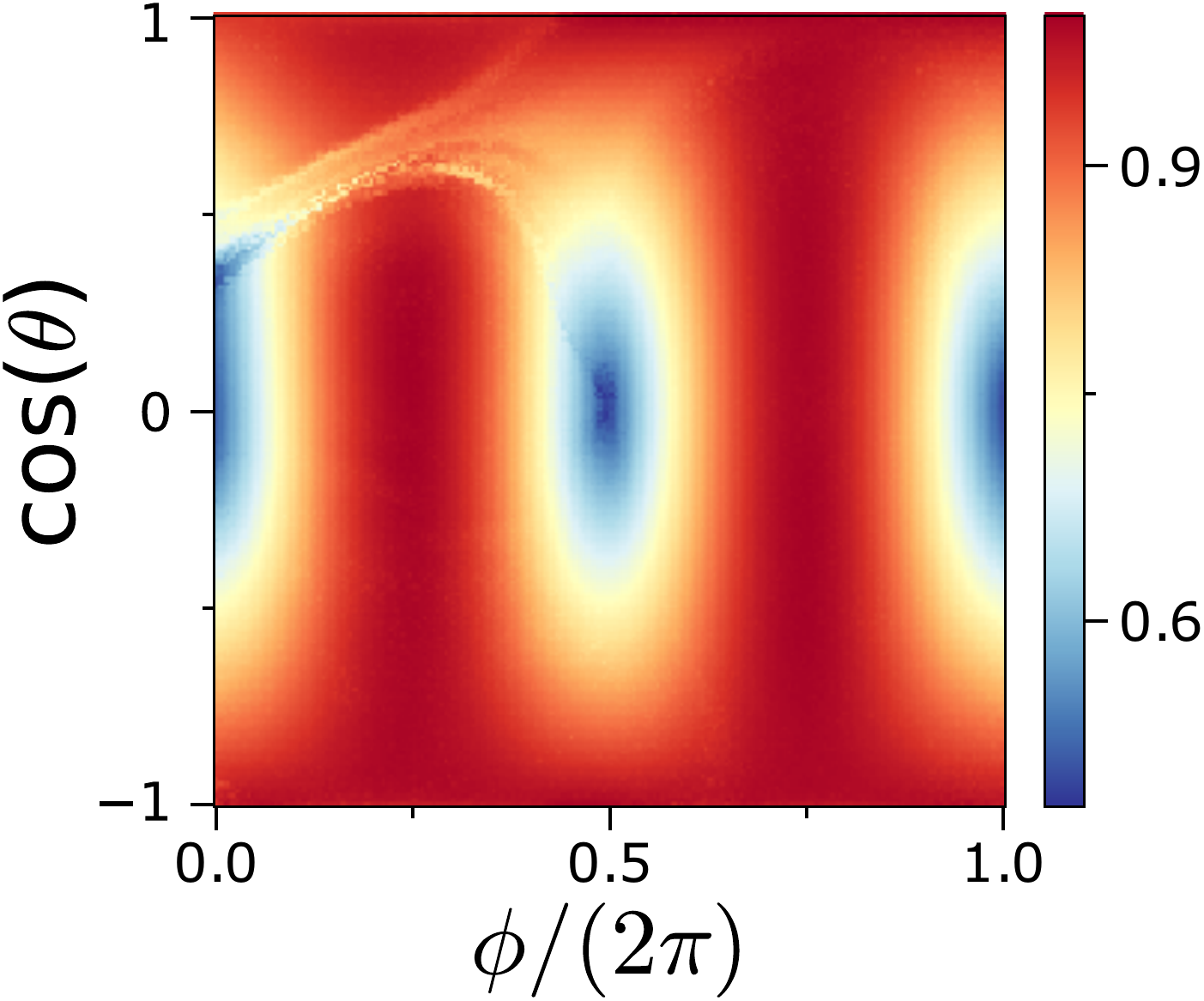}\hfill
	\subfigimg[width=0.24\textwidth]{b}{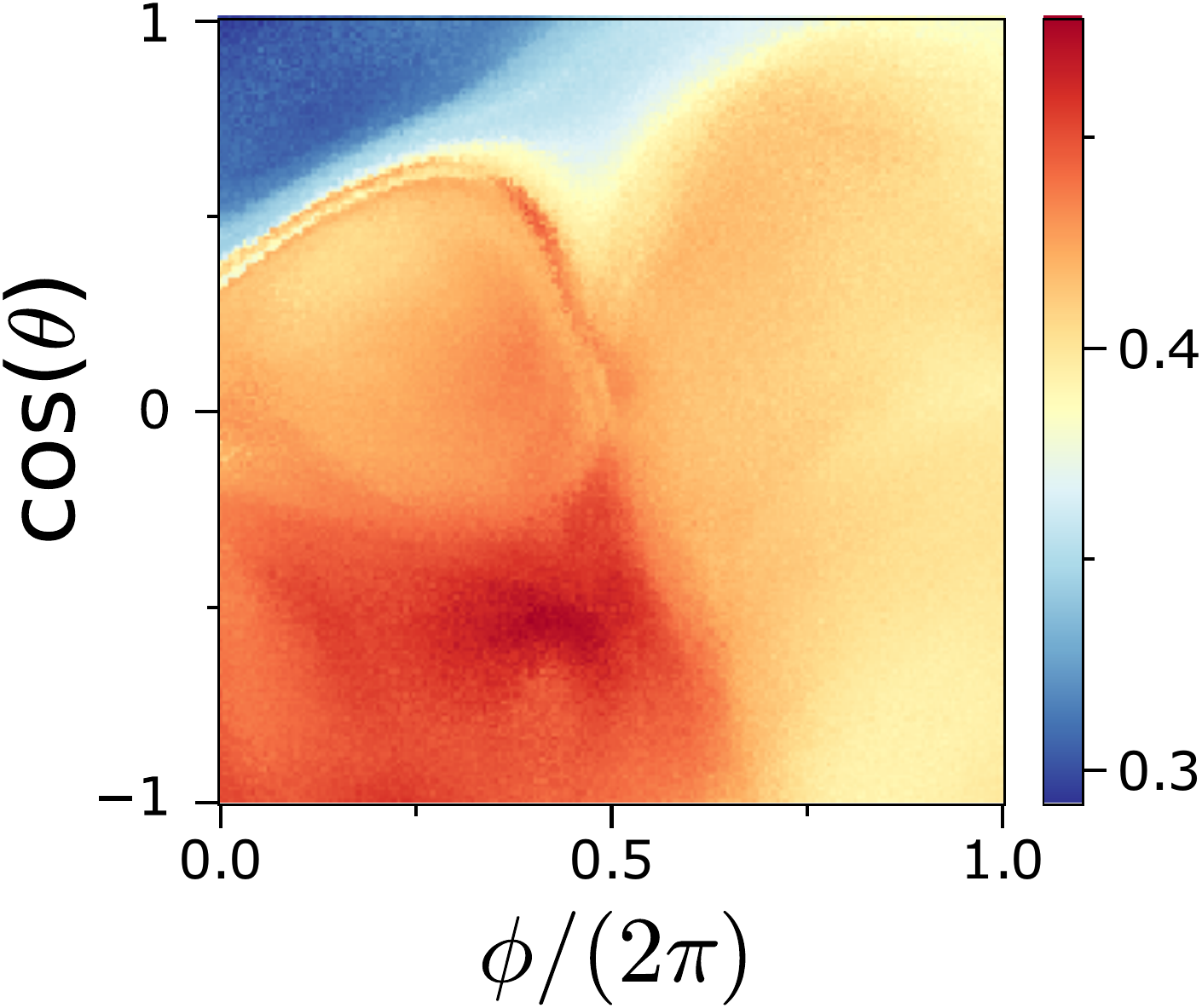}
	\caption{Create arbitrary quantum state in two-level subspace without applied magnetic field $B$ to break ground state degeneracy.  \idg{a} Fidelity of preparing state $\Psi(\theta,\phi)$ (Mean fidelity $\langle F\rangle=0.88$) \idg{b} Protocol time $T$.
		Parameters: $N_\text{T}=9$ timesteps, $n=600$ neurons, variable time per step with maximal time $0.2<T<0.8$, $-20<\Omega_{1,2}<20$, 800000 training epochs, detuning $\delta_1=50$, $\delta_2=0$ and $B_\text{ext}=0$. Using closed system Hamiltonian for the NV center.}
	\label{NVCenterB}
\end{figure}

The NV center has a degenerate ground state. The degeneracy can be broken by an external magnetic field. This is important to generate specific superposition states. As a comparison to the case of the main text, where a magnetic field is applied, we optimized the NV center without a magnetic field $B_\text{ext}=0$ in Fig.~\ref{NVCenterB}. We find that the fidelity for superposition states on the equator of the Bloch sphere with $\theta=\pi/2$ and $\phi=0$ or $\phi=\pi$ is decreased.

\section{Enforce periodic boundaries}
The angles for the target state $\theta$ and $\phi$ are given into the neural network to train it. If the angles are given as tuple of angles, the neural network does not understand that $\phi$ is actually a function with period of $2\pi$, e.g. $\phi=\phi \text{mod}2\pi$. The neural network has no knowledge of this and thus thinks $\phi=0$ is not the same as $\phi=2\pi$ as the numbers widely differ. For the neural network, two states are only close if they are close in value. This can be seen in Fig.2 as the protocol time $T$ indeed does not have the proper symmetry in $\phi$. 

\begin{figure*}[htbp]
	\centering
	\subfigimg[width=0.24\textwidth]{a}{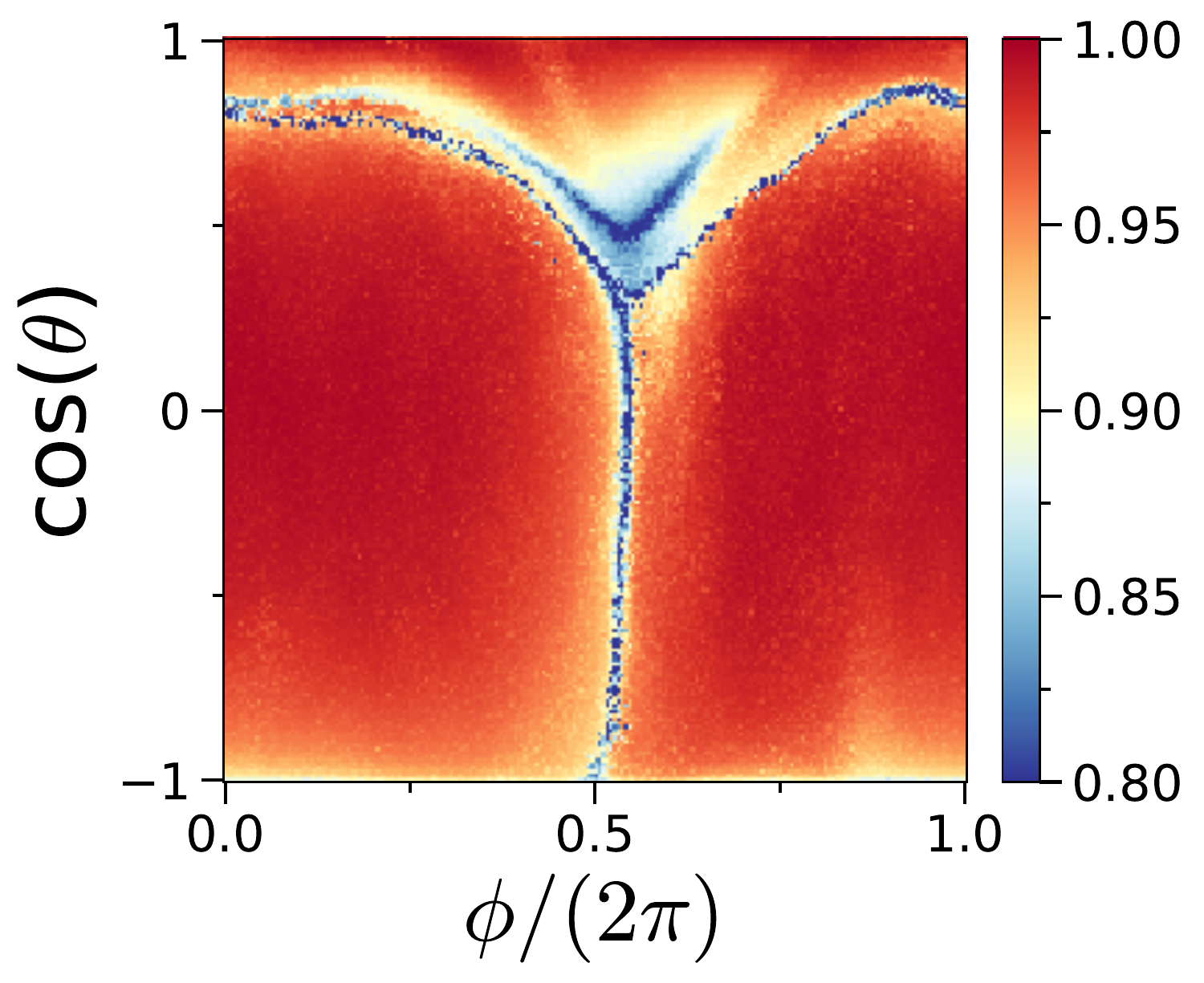}\hfill
	\subfigimg[width=0.24\textwidth]{b}{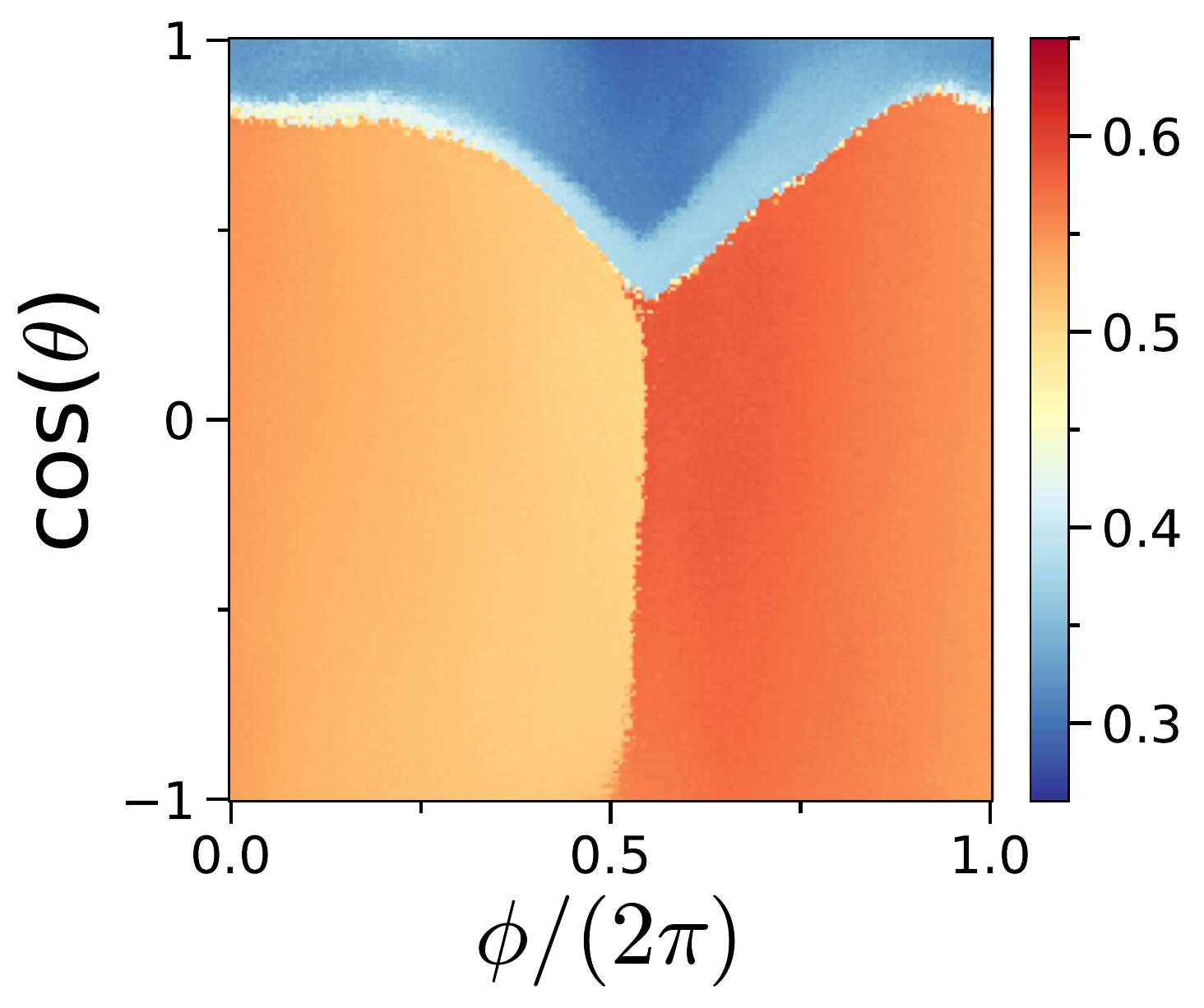}\hfill
	\subfigimg[width=0.24\textwidth]{c}{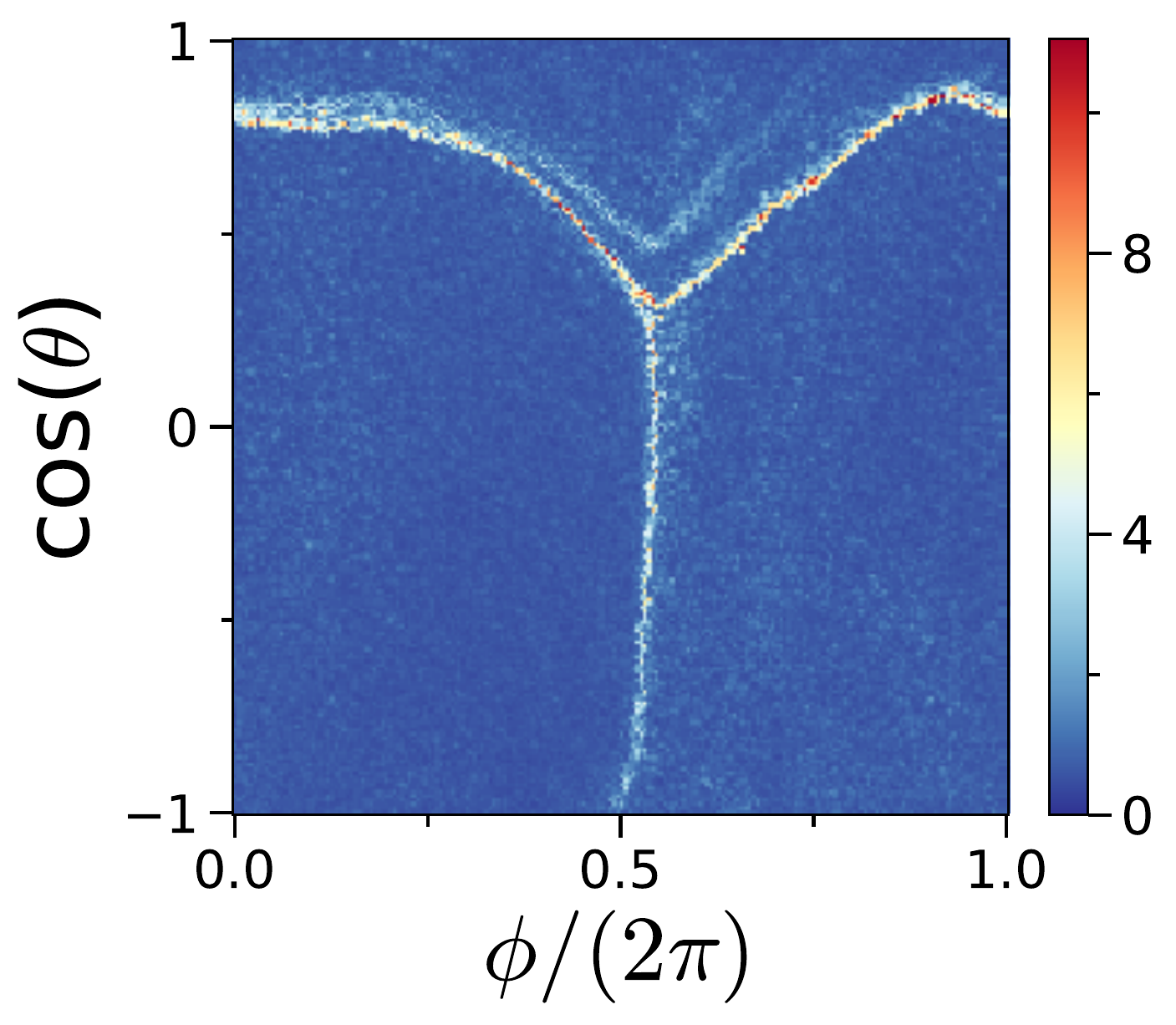}\hfill
	\subfigimg[width=0.27\textwidth]{d}{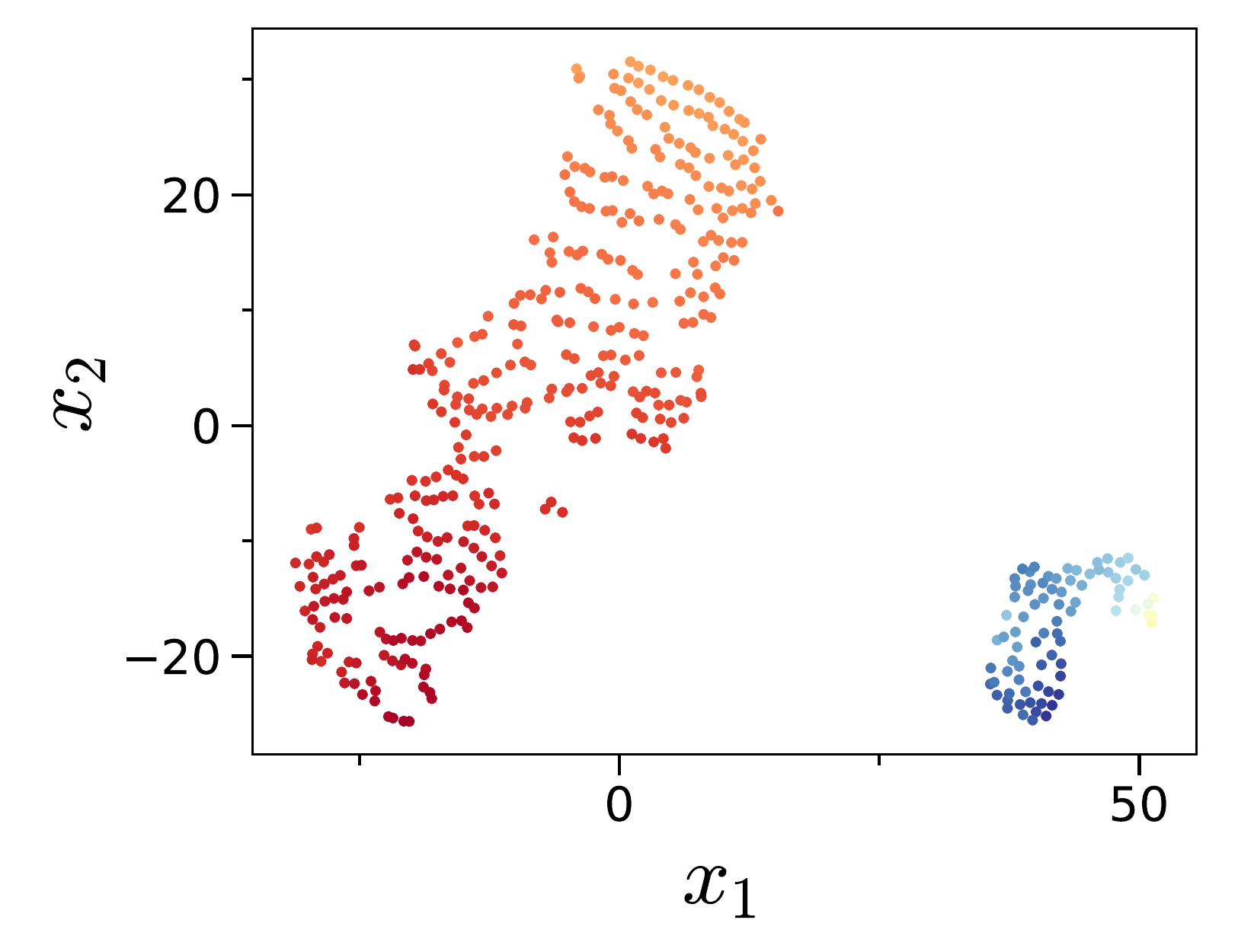}\hfill
	\caption{Create arbitrary quantum state $\Psi(\theta,\phi)$ with enforced periodic boundary conditions for $\phi$ (state parameterized for neural network by $\theta$, $\cos(\phi)$,  $\sin(\phi)$) using deep reinforcement learning.  \idg{a} Fidelity $F$ of preparing state  (Mean fidelity $\langle F \rangle=0.963$) \idg{b} Protocol gradient $G(\theta,\phi)$. Protocols are grouped into two distinct areas of low gradient, divided by a sharp lines of large gradient. Note that the system is now periodic in $\phi$. \idg{c} protocol time $T$. The areas are characterized by different protocol times $T$. The protocol time of the lower area changes with $\phi$, reflecting the impact of the magnetic field.
		\idg{d} Two-dimensional representation of the distance  between protocols using t-SNE algorithm. Color indicates protocol time. Protocols are close to each other if they are similar. 
		Parameters: $N_\text{T}=9$ timesteps, variable time per step with maximal time $0.2\text{ns}<T<0.8\text{ns}$, $-20\text{GHz}<\Omega_{1,2}<20$GHz, detuning $\delta_1=50$GHz, $\delta_2=0$ and $B_\text{ext}=0.15$T (all variables in units of $\hbar$), 650000 training epochs and $n=450$ neurons in two fully-connected layers.}
	\label{periodic}
\end{figure*}

The periodic boundary condition can be hard-coded by mapping $\phi$ to ($\cos(\phi)$, $\sin(\phi)$) and feeding those two numbers into the neural network. As this two numbers together have a periodic trajectory with $\phi$, the neural network understand that this is a periodic function. The result for this approach is shown in Fig.\ref{periodic}. The resulting graph has the proper symmetry. We observe that due to a different choice of input to the neural network, the resulting found near-optimal protocol landscape is different. This is not surprising, since there is a large number of nearly equivalent possible protocols for a problem with many control problems, such as the NV center. However, it seems enforcing periodic boundaries gives a slightly lower mean fidelity ($\langle F \rangle=0.963$ instead of $\langle F \rangle=0.972$).

\section{Spin rotations}
\begin{figure*}[htbp]
	\centering
	\subfigimg[width=0.8\textwidth]{}{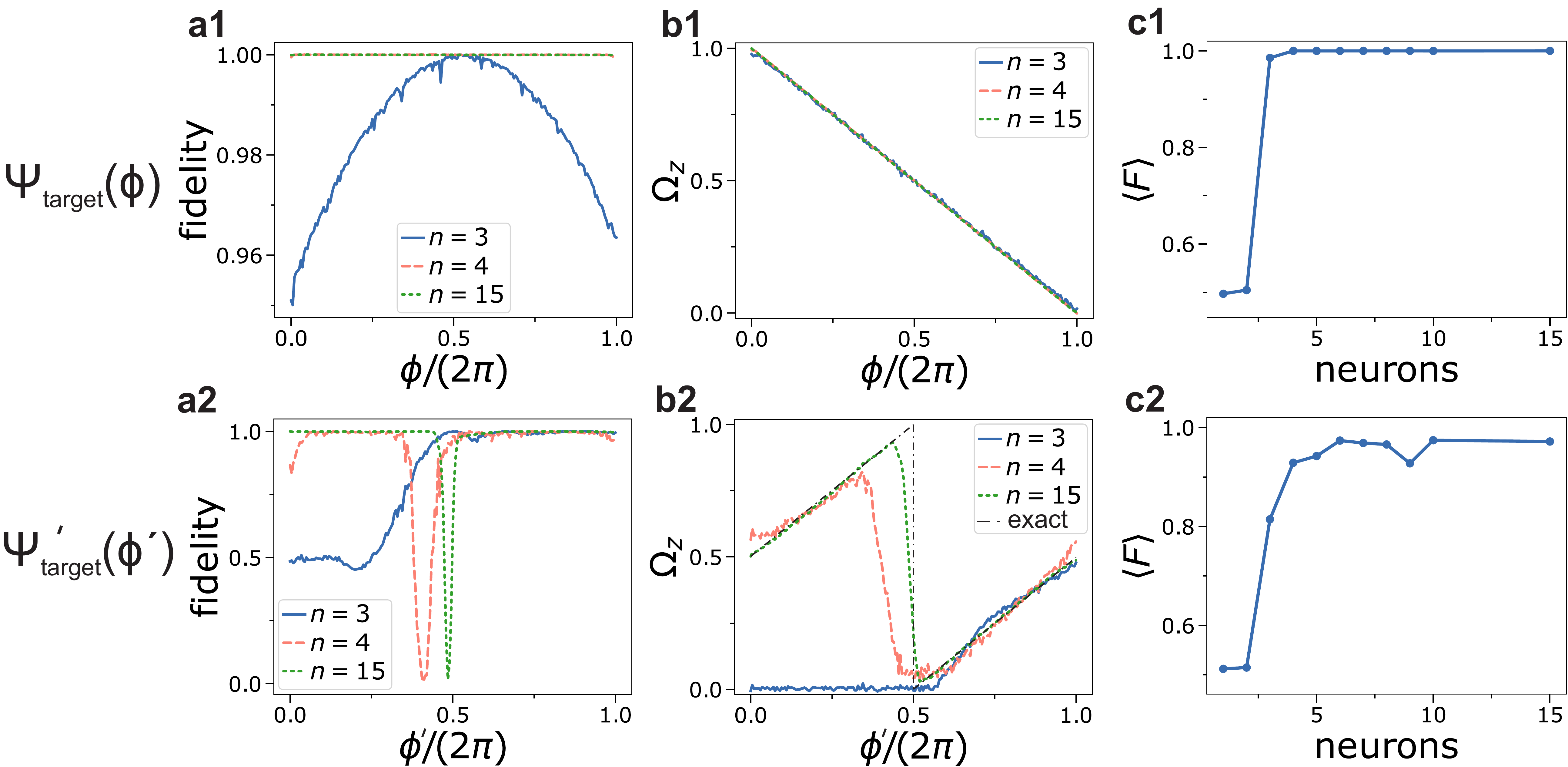}
	\caption{Learning preparing the state Eq.\ref{psip} on the equator of the Bloch sphere using two consecutive rotations around $y$ and $z$ axis respectively. We choose two different target states in the upper and lower row of graphs. Upper row: $\Psi(\phi,\theta=\pi/2)$ (Eq.\ref{psi}) Lower row: state $\Psi'(\phi',\theta=\pi/2)$ shifted by $\phi'=\phi+\pi$ (Eq.\ref{psip}). \idg{a} fidelity $F(\phi)=\abs{\braket{\Psi(\phi,\pi/2)}{{\Psi_\text{target}(\phi,\pi/2)}}}^2$ against phase of target state $\phi$ or $\phi'$ for different number of neurons of the neural network $n$. At least 3 neurons are needed to represent the driving protocol. 
		\idg{b} driving strength $\Omega_z$ for the rotation around the $z$-axis. 
		\idg{c} Fidelity $\langle F \rangle$ averaged over all $\phi$ ($\phi'$) for different number of neural network neurons.}
	\label{TwoLevel}
\end{figure*}
As a demonstration example, we consider a simple two-level system with basis states $\ket{0}$ and $\ket{1}$. We start with the initial state $\ket{0}$, then apply a rotation around the $y$-axis, followed by a rotation around the $z$-axis. We denote the angle of rotation $\Omega_y$ and $\Omega_z$, with the rotation unitaries $U_y(\Omega_y )=\exp\left(i\pi \Omega_y \sigma_y / 2\right)$ and $U_z(\Omega_z )=\exp\left(i\pi\Omega_z \sigma_z\right)$. To demonstrate our algorithm, we want to learn the rotation angles to generate arbitrary states on the equator of the Bloch sphere
\begin{equation}\label{psi}
\Psi(\phi,\theta)=\cos\left(\frac{\theta}{2}\right)\ket{0}+\sin\left(\frac{\theta}{2}\right)\expU{i\phi}\ket{1}\,. 
\end{equation} 
Furthermore, we constrain $\Omega_y$ and $\Omega_z$ to assume only values between 0 and 1. With this choice of constraints, every target state can be reached, as well as each target state corresponds to exactly one specific $\Omega_y$ and $\Omega_z$. $U_y$ can generate maximally a $\pi$ rotation around the $y$-axis, while $U_z$ maximally a full rotation around the $z$-axis.

The analytic solution is easy to write down: $\Omega_y(\phi)=\frac{1}{2}$, and $\Omega_z(\phi)=\frac{\phi}{2\pi}$.
The result of our learning protocol is shown in the upper row of Fig.~\ref{TwoLevel}.  With at least 3 neurons, our neural network can learn the optimal protocol with nearly unit fidelity for all possible target states.


Next, we shift the angle that parameterizes the target state. We re-define the target Bloch states by rotating them by $\pi$ around the $z$-axis: 
\begin{equation}\label{psip} \Psi'(\phi',\theta)=\cos\left(\frac{\theta}{2}\right)\ket{0}+\sin\left(\frac{\theta}{2}\right)\expU{i(\phi'-\pi)}\ket{1}\, . 
\end{equation}
Again, we constrain $\Omega_y$ and $\Omega_z$ assume only values between 0 and 1 such that every target state can be reached and each target state corresponds to exactly one specific $\Omega_y$ and $\Omega_z$. The protocols are shifted as well. The driving parameters to generate the shifted target state $\Psi(\phi')$ are $\Omega_y(\phi')=\frac{1}{2}$ and
\begin{equation}
\Omega_z(\phi')=
\begin{cases} 
\frac{\phi'+\pi}{2\pi} & 0\le\phi'\le\pi \\
\frac{\phi'-\pi  }{2\pi}& \pi< \phi'\le 2\pi
\end{cases}
\end{equation}
The optimal $\Omega_z(\phi')$ becomes now a discontinuous function. With this choice, we would like to demonstrate how the neural network learns discontinuous protocols compared to continuous ones.
The learning result is shown in the lower row of Fig.~\ref{TwoLevel}. We observe that even for increasing number of neurons, not all states can be perfectly created. There is a dip in protocol fidelity around $\phi'=\pi$. This is where the driving parameter $\Omega_z$ has to jump discontinuously, which is approximated as a finite slope in the neural network. As a result, a bad protocol is returned by the neural network in that region. The width of the dip decreases with increasing number of neurons. 

\begin{figure*}[htbp]
	\centering
	\subfigimg[width=0.24\textwidth]{a}{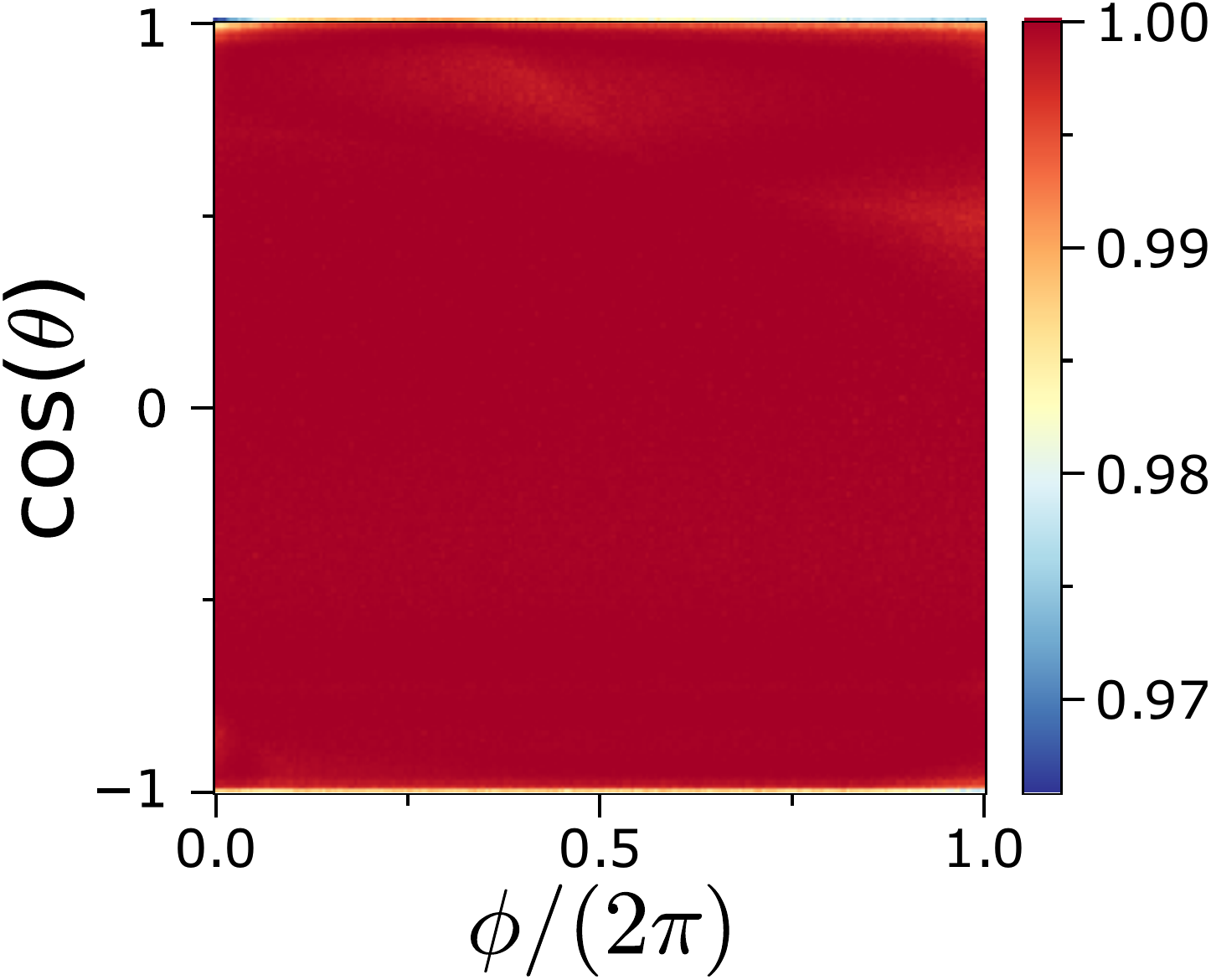}
	\subfigimg[width=0.24\textwidth]{b}{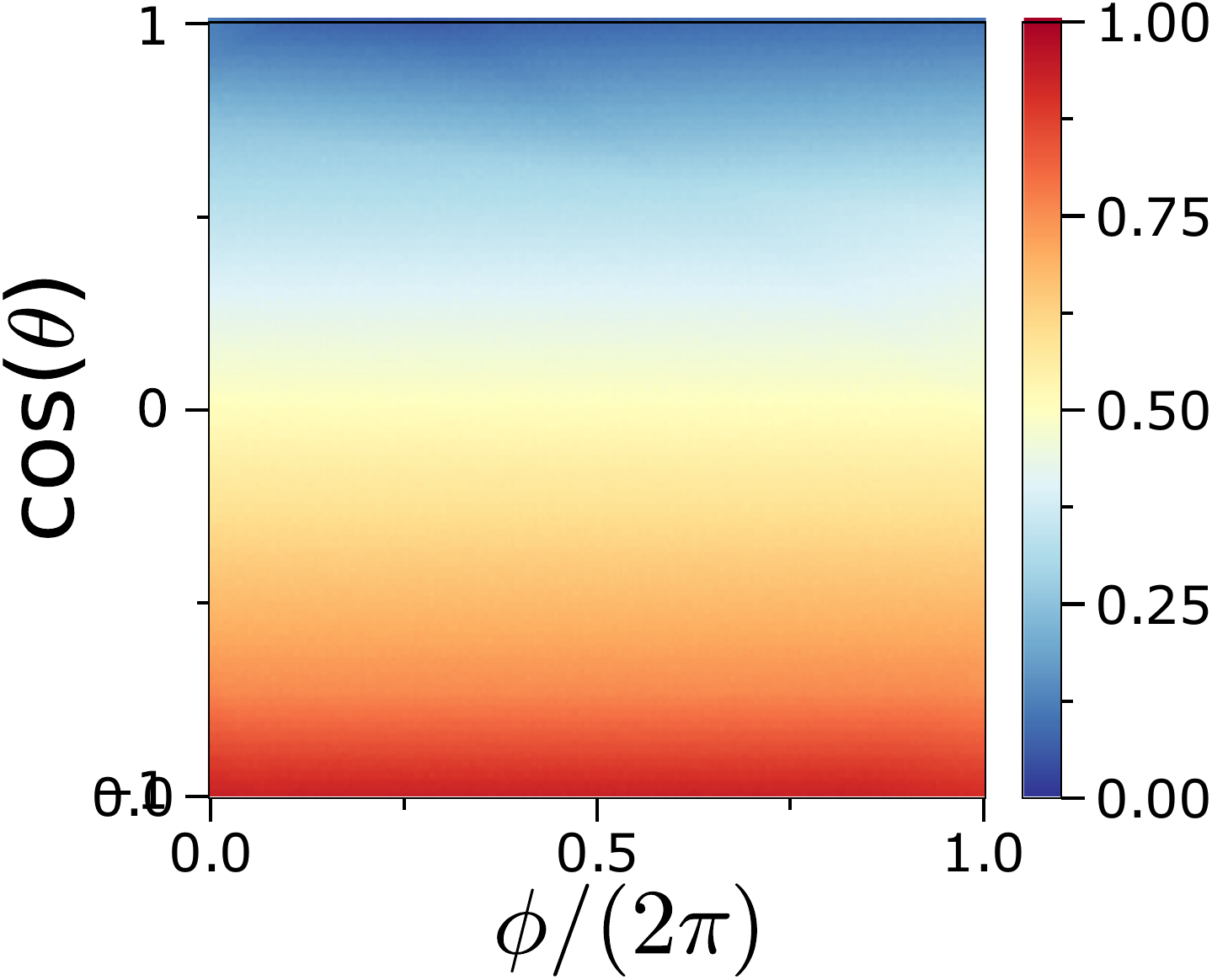}
	\subfigimg[width=0.24\textwidth]{c}{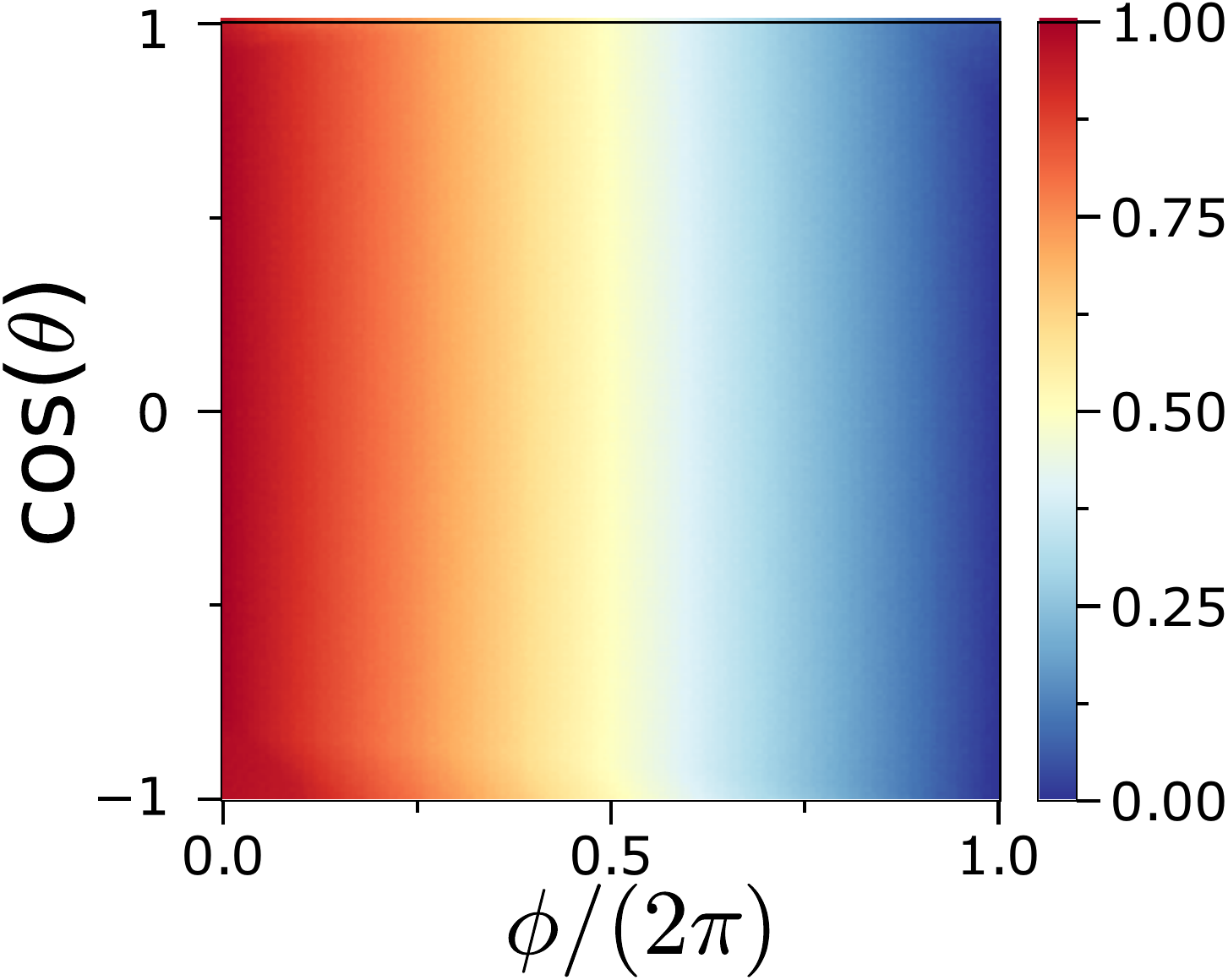}\\
	\subfigimg[width=0.24\textwidth]{d}{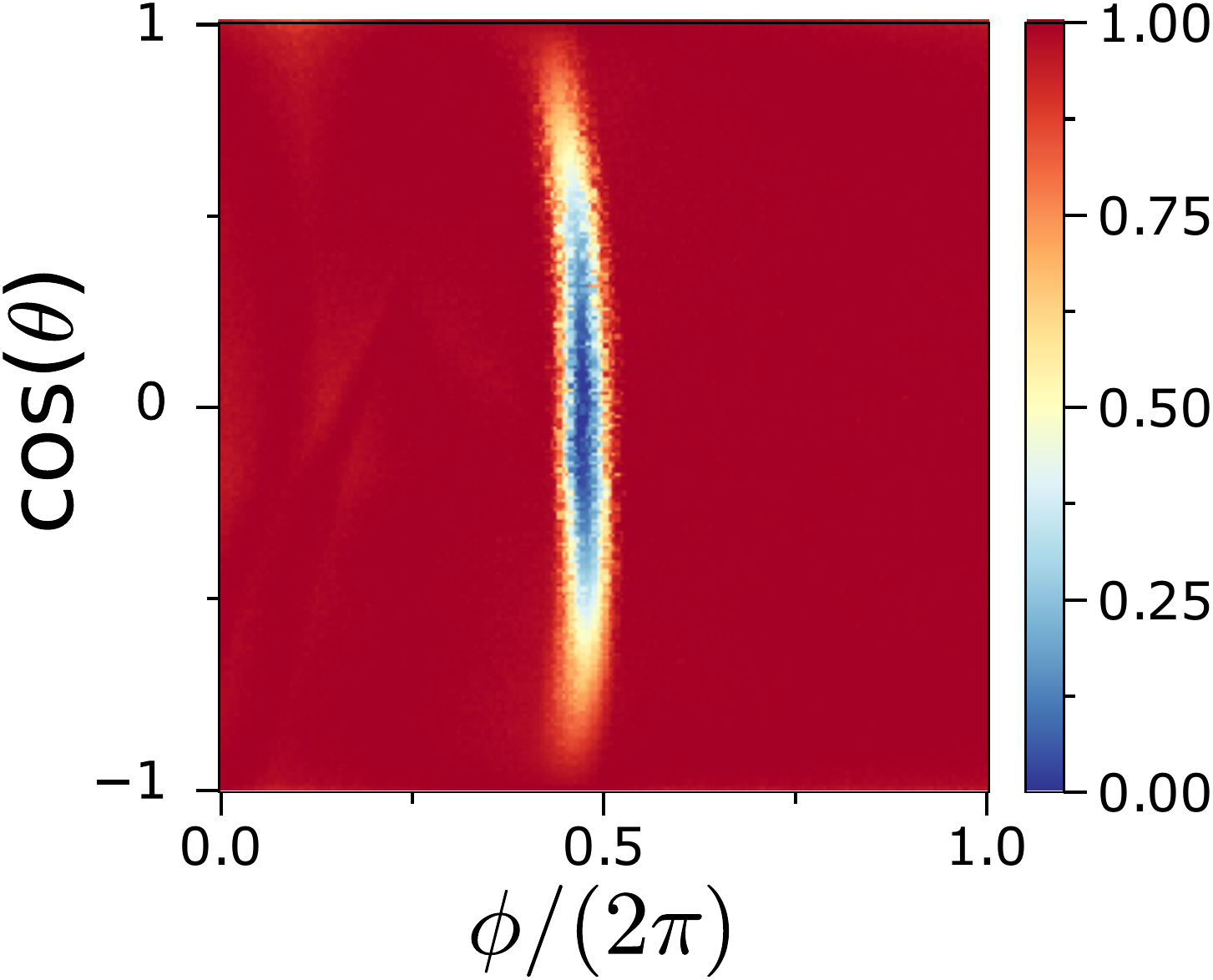}
	\subfigimg[width=0.24\textwidth]{e}{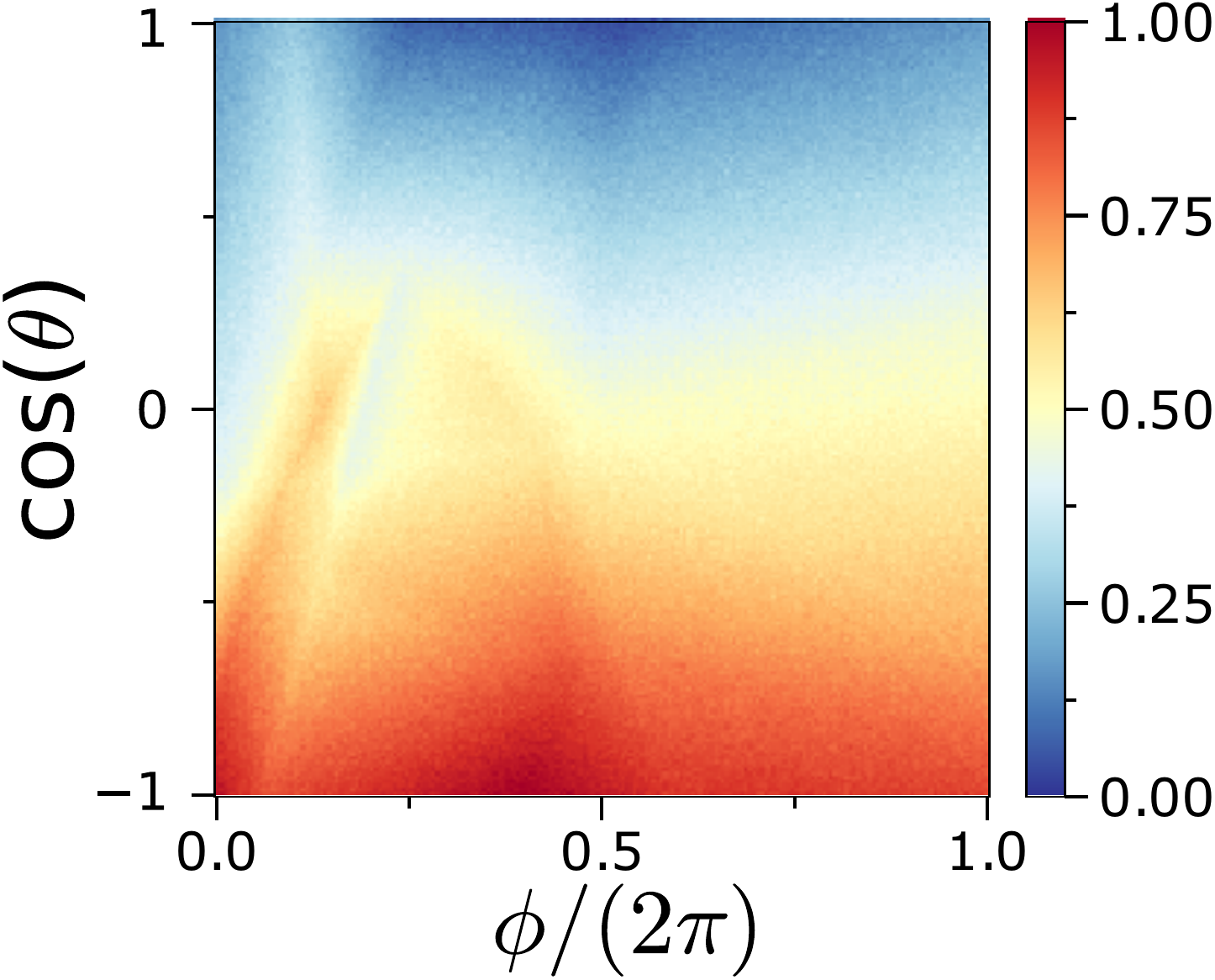}
	\subfigimg[width=0.24\textwidth]{f}{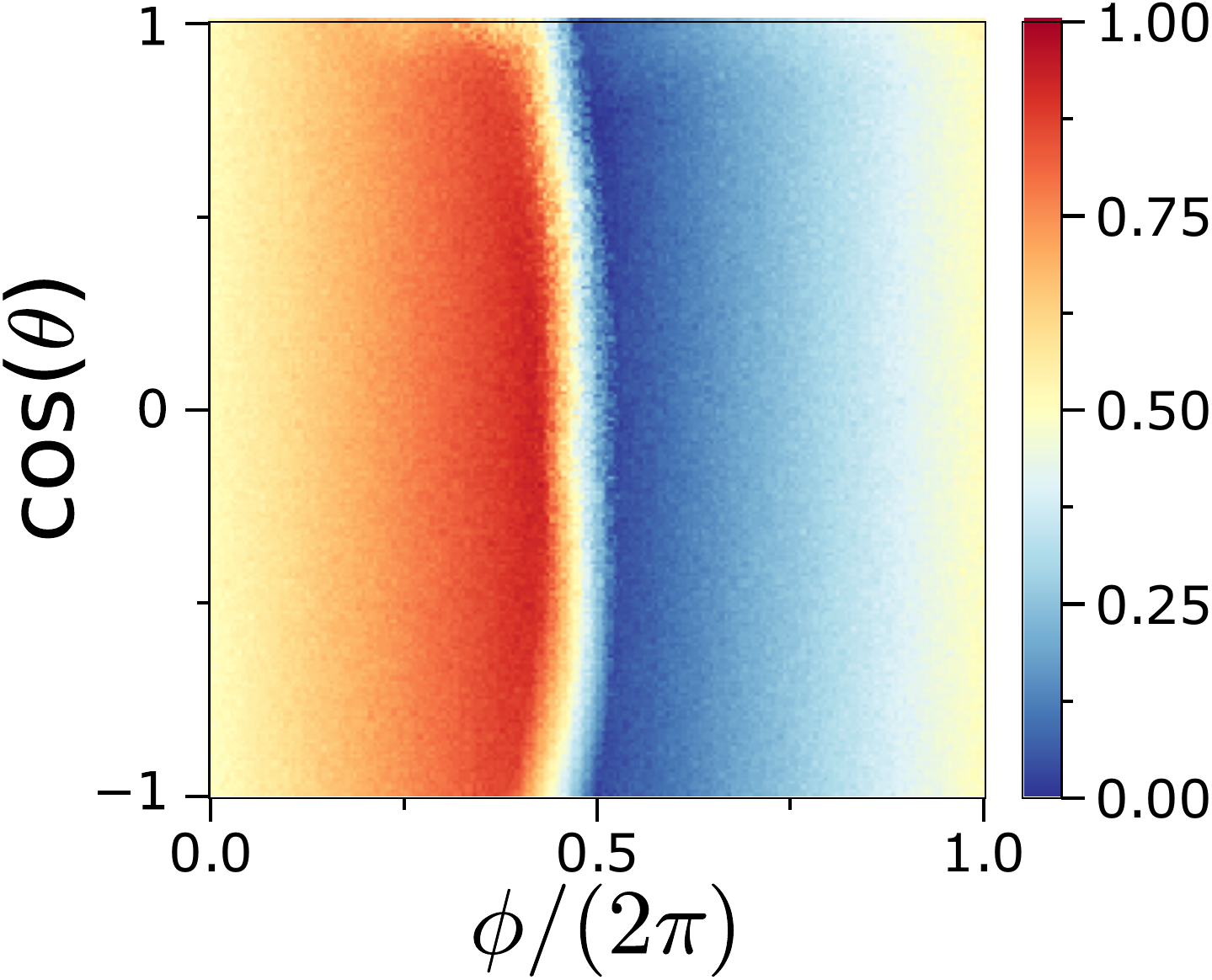}
	\caption{Learning preparing arbitrary states Eq.\ref{psi} (\idg{a,b,c}) and Eq.\ref{psip} (\idg{d,e,f}) on the Bloch sphere using two consecutive rotations around $y$ and $z$ axis respectively \idg{a,d} Fidelity for target state over angles $\theta$, $\phi$.  \idg{b,e} Driving protocol $\Omega_y$ for the rotation around the $y$-axis. \idg{c,f} Driving protocol $\Omega_z$ for the rotation around the $z$-axis. 
	}
	\label{TwoLevel2D}
\end{figure*}

The corresponding two-dimensional fidelity and driving protocol is shown in Fig.\ref{TwoLevel2D} for the cases of 10 neurons. Fig.\ref{TwoLevel2D}a,b,c shows the case for $\Psi(\phi,\theta)$ (Eq.\ref{psi}), and Fig.\ref{TwoLevel2D}d,e,f the case with shifted $\Psi'(\phi',\theta)$ (Eq.\ref{psip}). We plot fidelity in Fig.\ref{TwoLevel2D}a,d and the driving strength for $\Omega_y$ and $\Omega_z$ in Fig.\ref{TwoLevel2D}b,e and Fig.\ref{TwoLevel2D}c,f respectively. For the $\Psi(\phi,\theta)$ case we observe a clean and smooth variation of the protocol and high fidelity for all $\theta$ and $\phi$.
For $\Psi'(\phi')$ we observe that around $\phi\approx\pi$ the fidelity decreases to zero (Fig.\ref{TwoLevel2D}d) and the driving strength for $\Omega_z$ (Fig.\ref{TwoLevel2D}f) varies rapidly. Here, the neural network tries to interpolate between the two areas of $\phi<\pi$ and $\phi>\pi$, which requires a sudden jump in the driving protocol. As the neural network can only insufficiency approximate this step function, the fidelity decreases drastically.

The representation power of the neural network (e.g. the complexity of functions that the neural network can generate) increases with number of neurons. For a low number of neurons, it cannot represent large gradients in the protocol landscape. 
A sudden dip in the fidelity is a indication of a sudden change in the protocol landscape.

\end{document}